\documentclass{aa}  
\usepackage{graphicx}
\usepackage{txfonts}

\usepackage{placeins}

\begin{document}

  \title{How the origin of stars in the Galaxy impacts the composition of planetary building blocks}

   \author{N. Cabral \inst{1},  A. Guilbert-Lepoutre \inst{1},  B. Bitsch\inst{2}, N. Lagarde \inst{3}  and S. Diakite \inst{4}}

   \institute{Laboratoire de G\'eologie de Lyon, CNRS UMR5276, Univ. Claude Bernard Lyon 1, ENS Lyon, F-69622, Villeurbanne, France\\
              \email{nahuel.cabral@univ-lyon1.fr}
      \and
              Max-Planck-Institut f\"ur Astronomie, K\"onigstuhl 17, 69117 Heidelberg, Germany
      \and
             Laboratoire d’Astrophysique de Bordeaux, Université Bordeaux, CNRS, B18N, Allée Geoffroy Saint-Hilaire, 33615 Pessac, France
        \and
        M\'{e}socentre de Franche-Comt\'{e}, Universit\'{e} de Franche-Comt\'{e}, 16 route de Gray, 25030 Besan\c{c}on Cedex, France
        }

   \date{Received  xx xx xx / Accepted 29 November 2022}

   \titlerunning{How the origin of stars in the Galaxy impacts the composition of planetary building blocks}
   
   \authorrunning{Cabral et al.}

  \abstract 
      {Our Galaxy is composed of different stellar populations with varying chemical abundances, which are thought to imprint the composition of planet building blocks (PBBs). As such, the properties of stars should affect the properties of planets and small bodies formed in their systems. In this context, high-resolution spectroscopic surveys open a window into the chemical links between and their host stars.}
   {We aim to determine the PBB composition trends for various stellar populations across the Galaxy by comparing the two large spectroscopic surveys APOGEE and GALAH. We assess the reliability of the PBB composition as determined with these surveys with a propagation error study.} 
   {Stellar spectroscopic abundances from the large surveys GALAH-DR3 and APOGEE-DR17 were used as input with a stoichiometric condensation model. We classified stars into different Galactic components and we quantified the PBB composition trends as a function of [Fe/H]. We also analysed the distribution composition patterns in the [$\alpha$/Fe]-[Fe/H] diagram.}
   {Our propagation error study suggests that the overall trends with [Fe/H] and [$\alpha$/Fe] are robust, which is supported by the double study of both APOGEE and GALAH. We therefore confirm the existence of a bimodal PBB composition separating the thin disc stars from the thick disc stars. Furthermore, we confirm that the stoichiometric water PBB content is anti-correlated with [Fe/H].} 
   {Our results imply that metal-poor stars both in the thin and thick disks are suitable hosts for water-rich PBBs and for ice-rich small bodies. However, for metal-poor stars ([Fe/H]<0), the PBBs around thick disc stars should have a higher water content than that around thin disc stars because of the $\alpha$-content dependence of the water mass fraction. Given the importance of the initial water abundance of the PBBs in recent planet formation simulations, we expect that the star origin influences the exoplanet population properties across the Galaxy.}
   
   \keywords{Galaxy: stellar content; Stars: abundances; Stars: statistics; Stars: Planetary systems; Comets: general}

   \maketitle

\section{Introduction}

The properties of exoplanets appear to correlate with the chemical properties of their host stars. It has been observed that stars with increasing [Fe/H] ratio harbor more giant planets \citep[][]{Santos2004, FischerValenti2005, Johnson2010}. Small planets orbiting metal-poor stars present larger periods \citep{BeaugeNesvorny2013, Wilson2018}, and giant planets seem to have lower eccentricities when orbiting metal-poor stars \citep{DawsonMurray-Clay2013,Buchhave2018}. In addition, iron-poor stars hosting planets are found to preferentially present an enhanced alpha-element composition \citep[e.g.][]{Haywood2008,Haywood2009,Adibekyan2012a,Adibekyan2012b}. Interestingly, the occurrence of small planets appears related to stellar population properties \cite[e.g.][]{BashiZucker2019, Bashi2020, BashiZucker2022}, and it could be related to the chemical environment in which they were formed. More generally, it is reasonable to expect that the native environment of planetesimals and planets should impact their properties. In particular, their bulk composition could reflect the chemical properties of their host star. Indeed, observational constraints tend to show a high correlation between the exoplanet and the host star chemical abundances \citep{Adibekyan2021}. Across the Galaxy, the different stellar populations are thought to produce planet building blocks (PBBs) \citep{Santos2017,Cabral2019} and planets \citep{Bitsch2020} with different compositions. It is thus relevant to consider that the chemical properties of host stars are important in the context of planet formation models. However, the chemical links between stars and bodies in their planetary systems may be difficult to disentangle, because of the diversity of physical and thermal processes involved in the formation processes of these objects.

One approach is to compute the composition of PBBs formed when solids condense from the gaseous disc \citep[][]{Santos2017,Cabral2019,Bitsch2020}. This method is particularly useful when analysing general trends for stellar populations in our Galaxy. In this context, understanding the link between stellar and PBB compositions is crucial to understanding how stellar populations impact planet properties. 
To reach a statistically significant sample of stars representative of the diversity of the stellar populations of the Galaxy, the study of large spectroscopic surveys is required.  Over the last decade, a significant effort has been devoted to the development of large spectroscopic surveys; for instance, RAVE  \citep{Steinmetz2020a,Steinmetz2020b},  SEGUE  \citep{Yanny2009},  Gaia-ESO \citep{Recio-Blanco2014},  APOGEE \citep{Abdurro'uf2021},  LAMOST \citep{Zhao2012}, and GALAH \citep{Duong2018}. Some of them provide high-resolution spectra, substantially increasing the quality of chemical abundance determination. Based on such observational data, numerous studies have underlined the existence of a gap between the thin and the thick discs \citep{Recio-Blanco2014, Hayden2015, Duong2018}, supporting the original idea of \cite{Yoshii1982} and \cite{GilmoreReid1983}. In this framework, it is generally thought that stellar populations in the Milky Way, which display different metallicities and $\alpha$ abundances, are the result of different formation mechanisms, epochs, and different chemical evolutions. The Galactic disc exhibits two sequences, where thick-disc stars are generally metal-poor and alpha-enriched when compared to thin disc stars \citep[e.g.][]{Haywood2013, Kordopatis2015}. The halo contains the more metal-poor stars \citep[e.g.][and references therein]{FernandezAlver2018}. The metallicity range of the bulge is similar to that of the thin disc, but the spread in alpha abundances is far larger \citep[e.g.][]{BarbuyChiappini2018,RojasArriagada2019}. 

\citet[][hereafter S17]{Santos2017} computed the PBB composition using 371 HARPS stars. They found different chemical composition between the thin disc and the thick disc. In particular, the thin disc presents iron and water mass fractions that are, respectively, higher and lower than the thick disc. With synthetic simulations based on the Besançon Galaxy Model \citep{Lagarde2021}, \citet[][hereafter C19]{Cabral2019} studied the PBB composition in the [$\alpha$/Fe]-[Fe/H] diagram. They simulated the chemistry of millions of synthetic stars in order to study the potential link between the PBB composition and the stellar populations across the Galaxy: thick disc, thin disc, halo, and bulge. In particular, they found that the well-known stellar density gap in the [$\alpha$/Fe]-[Fe/H] diagram between the thin and the thick discs results in a bimodal distribution of PBB composition (see their histograms in Fig. 2 and Fig. 3). This suggests that the chemical composition in the early phases of proto-planetary discs could greatly differ depending on the galactic origin of the host star.

In this work, we aim to study the PBB composition patterns in the [$\alpha$/Fe]-[Fe/H] diagram. The goal is to determine how the chemical specificities of the thin and the thick discs can impact the final PBB composition. We took advantage of the significant statistics and the increasing accuracy in the observed stellar abundances to constrain the expected PBB composition. The latest releases of the large spectroscopic surveys APOGEE DR17 and GALAH DR3 offer excellent data to analyse and compare. In addition, we want to assess the robustness of the stoichiometric predictions with respect to the typical errors in spectroscopic abundance determinations. For this, we computed a simple propagation error test to determine how much the PBB composition is modified when taking into account spectroscopic error bars. This work thus continues the study of C19, with the updated stoichiometric model used by \citet[][hereafter BBB20]{Bitsch2020}, and makes use of the exceptional observational context of the large, high-resolution surveys APOGEE-DR17 and GALAH-DR3. In Section \ref{Method}, we describe the selected stellar samples, the galactic classification methods, and the stoichiometric model we applied. Section \ref{SecMassFractionMetallicity} determines the PBB composition as a function of the metallicity, [Fe/H], and compares the results with BBB20. Section \ref{SecDiag} shows the PBB composition in the [$\alpha$/Fe]-[Fe/H] plane and discusses the thin/thick disc differences. Section \ref{SecErrors} presents a propagation error study to evaluate the reliability of PBB composition values. Finally, in Sections 6 \label{SecRadius} and \ref{Conclu} we draw our conclusions.

\section{Methods}
\label{Method}

\subsection{Stellar sample} 
\label{SecSample}

In order to have a broad picture of the expected chemical compositions of the different stellar populations, we used the large spectroscopic surveys GALAH and APOGEE. For the purpose of this study, we analysed both surveys simultaneously but separately, because the determination of stellar compositions is not necessarily derived in a homogeneous way. This approach enables a simple comparison and it allows us to discuss the robustness of resulting trends.

We impose a series of quality cuts to the full APOGEE-DR17 and GALAH-DR3 releases. We first require a signal-to-noise of S/N>100 for APOGEE-DR17 and SNR>30 for GALAH-DR3 as providing a compromise between the number of analysed stars and the potential bias on the overall trends. In addition, for APOGEE we used the following parameter quality flags (equal to 0): STARFLAG, ANDFLAG, FE$\_$H$\_$FLAG, and we removed stars with the STAR$\_$BAD and STAR$\_$WARN flags. For GALAH, we selected stars with the following parameter quality flags (equal to 0): flag$\_$sp and flag$\_$fe$\_$h. Both surveys use good elemental abundance quality flags (i.e. equal to zero) for S, Si, Mg, C, and O. We required each selected star to have all those quality flags. When computing [$\alpha$/Fe]\footnote{The alpha content is computed here with \([\alpha/Fe] = ([Mg/Fe] + [Si/Fe] + [Ca/Fe] + [Ti/Fe])/4\).} (cf. Sect. \ref{SecDiag}), we also required good-quality flags for Ca and Ti for every star. 

Stellar abundances from the early stellar phases are taken into account in priority, because the pre- and main sequence should be, a priori, more representative of the original abundances in the proto-planetary disc than abundances observed in evolved stars. We select pre-main and main-sequence stars with a simple and standard criterion: \( \text{log} \,g<4 \) and \( T_\text{eff} < 6400 K \). Moreover, our selected sample does not include stars with $T_\text{eff}$<4500 K.

Moreover, since there are no S-abundance determinations in the GALAH survey, we assume that it scales as Si. This trend has been shown in \cite{Chen2002} and confirmed in several studies 
\citep{Caffau2005, Jonsson2011, Takeda2016, Duffau2017}. We note that this trend is also consistent with our APOGEE selected sample.

\subsection{The [$\alpha$/Fe]-[Fe/H] plane} 
\label{SecSampleDiscsPhases}

One goal of this study is to analyse the PBB composition in the [$\alpha$/Fe]-[Fe/H] plane. Usually, the observed double sequence of the Milky Way discs observed in the [$\alpha$/Fe] versus [Fe/H] plane is associated with the chemical thin and thick discs at the solar circle. Recent studies underlined the existence of a gap between the thin and thick discs \citep{Fuhrmann2004,Reddy2006,Bensby2007}. These findings have been confirmed with higher spectral resolution (HARPS: \cite{Adibekyan2013}; Gaia-ESO: \cite{Recio-Blanco2014}; GALAH: \cite{Duong2018}; APOGEE-DR10: \cite{Anders2014}; APOGEE-DR12: \cite{Hayden2015}; APOGEE-DR16: \cite{Queiroz2020}; APOGEE-DR17: \cite{Abdurro'uf2021}). In particular, the thin disc stars are alpha-poor and tend to be metal-rich, while the thick disc stars are alpha-rich and tend to be metal-poor. This leads to a bimodal density distribution in the [$\alpha$/Fe]-[Fe/H] plane.

The presence of a third stellar population is still in debate. Based on the density distribution in the diagram [$\alpha$/Fe]-[Fe/H] (cf. their Fig. 6), \cite{Lagarde2021} considered two thick disc populations: the high-alpha, metal-rich thick disc and the high-alpha, metal-poor thick disc \citep[see also][]{Adibekyan2011}. Interestingly, the high-alpha, metal-rich thick disc has kinematics properties closer to the thin disk than the high-alpha, metal-poor thick disc. However, in this work we restrict our classification to the classical thin disc and thick disc.

\begin{figure}
\centering
\includegraphics[width=8cm,clip=true, trim= 0cm 0cm 0cm 0cm]{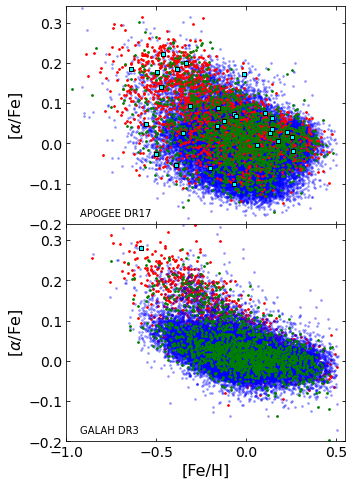}
\caption{Distribution of stars in the [$\alpha$/Fe] versus [Fe/H] plane. Top panel shows the APOGEE-DR17 sample while the bottom panel shows the GALAH-DR3 sample. The kinematical classification corresponds to thin disc stars (blue), thick disc stars (red), intermediate populations (green), and halo stars (cyan squares).}
\label{Fig_Density}
\end{figure}

\subsection{Classification of Galactic components}
\label{secClassGal}

We aim to classify our selected stars into the three Galactic components: thin disc, thick disc, intermediate members, and halo members. However, we recall that there is no obvious method to obtain a sample of a purely single Galactic component. Any method will produce samples contaminated by the other Galactic components, because the thin and the thick discs overlap in their spatial, kinematical, and chemical distributions.

The bimodal density distribution in the [$\alpha$/Fe]-[Fe/H] plane is widely used to separate thin and thick disc stars \citep{Adibekyan2013, Bensby2014,Lagarde2021}. However, the separation lines between the high-$\alpha$ sequence and the low-$\alpha$ sequence can differ from author to author because of potential differences in the observational surveys and the selection sample. We only used this method (Appendix \ref{AppendixDensityPlot}) as a comparison to the kinematical classification method, which is our nominal methodology to classify stars in this work.

In the kinematical classification approach, the different Galactic components (thin disc, thick disc, intermediate population and halo) are based on their Galactic positions and velocities by adopting the widely used kinematic approaches from \cite{Bensby2003, Bensby2014}. This probabilistic approach assumes the Galactic velocities of the LSR ($U_{LSR}$, $V_{LSR}$, $W_{LSR}$) have multi-dimensional Gaussian distributions:

\begin{equation}
        P = k \times exp  \left(         - \cfrac{U^2_{LSR}}{2{\sigma_U}^2}  -  \cfrac{(V_{LSR}-V_{asym})^2}{2{\sigma_V}^2}   - \cfrac{W^2_{LSR}}{2{\sigma_W}^2  }                              \right)
,\end{equation}

where $\sigma_U$, $\sigma_V,$ and $\sigma_W$ are the characteristic velocity dispersions and $V_{asym}$ and $U_{asym}$ are the asymmetric drifts. The normalisation coefficient is defined by

\begin{equation}
        k =  \cfrac{1}{ (2\pi)^{3/2}  \sigma_U \sigma_V \sigma_W   } 
.\end{equation}

The relative probabilities between two different components -TD/D (thick-disc to thin-disc), TD/H (thick disc to halo)- can be calculated as follows:

\begin{equation}
        \cfrac{TD}{D} = \cfrac{X_{TD}}{X_D}  \cdot  \cfrac{P_{TD}}{P_D}  \\
        \cfrac{TD}{H} = \cfrac{X_{TD}}{X_H}  \cdot  \cfrac{P_{TD}}{P_H} 
,\end{equation}

 where X is the fraction of stars for a given galactic component. The probability of belonging to one of the components has to be significantly higher than the probability of belonging to the others, to assign a target to it \citep{Bensby2014}.

The main kinematic parameters (U, V, W) for the stars were calculated using the Python-based package for galactic-dynamics calculations \textit{galpy}\footnote{http://github.com/jobovy/galpy} by \cite{Bovy2015}. The proper motions, coordinates, and radial velocities were taken from the Gaia data release 3 catalogue \citep{GaiaCollab2016b, GaiaCollabEDR32021, Lindegren2021,Seabroke2021}.
As expected from a kinematical perspective, Fig. \ref{Fig_Density} shows that the thick disc stars (red points) are more spread out on the alpha axes, while thin disc stars (blue points) are concentrated in the low-alpha zone.


\subsection{Chemical model}
\label{SecChemicalModel}

In proto-planetary discs, the bulk mineralogy of PBB is controlled by the ratios of Mg/Si and C/O. Under the assumption of equilibrium, proto-planetary discs with C/O>0.8 will contain carbon-rich phases (such as graphite, SiC, and TiC), only the outer part of the proto-planetary disc will have olivine and pyroxene \citep{Bond2010}.

For C/O<0.8, Si will exist in the solid form as SiO$_4$ or SiO$_2$, predominantly forming Mg-silicates. In this case, there are three regimes of mineral formation: (1) when Mg/Si<1, the magnesium primarily forms pyroxene (MgSiO$_3$) and the remainder of the silicon forms feldspars or olivine (Mg$_2$SiO$_4$); (2) when 1<Mg/Si<2, there is mixture of pyroxene and olivine similar to that of the Solar System; (3) when Mg/Si>2, silicon forms olivine, and the remainder of the magnesium will form magnesium compounds
such as MgO and MgS under specific (T, P) conditions \citep{Bond2012}.

We used stoichiometric relations from BB20. Their calculation of the water mass fraction accounts for CO, CO$_2,$ and CH$_4$. The gaseous molecules of CO and CO$_2$ bind many oxygen atoms that are not available to be condensed in water ice. These stoichiometric relations consider the case of 1<Mg/Si<2, which actually accounts for most of the observed stars. However, the third release of the GALAH survey extended the number of stars. In our selected samples, we found approximately 10\% of GALAH-DR3 stars and 4\% of APOGEE-DR17 stars with Mg/Si<1. We note that with our selection criteria, the number of stars with Mg/Si<1 is negligible in GALAH-DR2 consistently with BB20. Moreover, we have  less than 10\% of GALAH-DR3 stars and a totally negligible amount of APOGEE-DR17 stars with Mg/Si>2. Consequently, by simplicity we exclude stars with Mg/Si>2 but we account for stars with Mg/Si$\leqslant$1 with the following stoichiometric relations:

\begin{equation}
\begin{split}
& \;\;\;\;\;\;\;        N_{MgSiO_{3}} = N_{Mg} \\
& \;\;\;\;\;\;\;        N_{SiO_{2}} = N_{Si} - N_{Mg} \\
& \;\;\;\;\;\;\;        N_{FeS} =  N_{S} \\
& \;\;\;\;\;\;\;        N_{Fe_{2}O_{3}} = 0.25 \times (N_{Fe} - N_{S}) \\
& \;\;\;\;\;\;\;        N_{Fe_{3}O_{4}} =  (1/6) \times (N_{Fe} - N_{S}) \\
& \;\;\;\;\;\;\;        N_{CO} = 0.45 \times N_{C} \\
& \;\;\;\;\;\;\;        N_{CH_{4}} = 0.45 \times N_{C} \\
& \;\;\;\;\;\;\;        N_{CO_{2}} = 0.10 \times N_{C} \\
& \;\;\;\;\;\;\;        N_{H_{2}O} = N_{O} - (3 \times N_{MgSiO_{3}} + 2 \times N_{SiO_{2}} + N_{CO} \\ 
& \;\;\;\;\;\;\;        + 2 \times N_{CO_{2}} + 3 \times N_{Fe_{2}O_{3}} + 4 \times N_{Fe_{3}O_{4}}) \\
\end{split}
,\end{equation}

where, N$_X$ represents the number of each species, X, relatively to hydrogen. Despite the non-negligible proportion of stars with Mg/Si<1, in this study the SiO$_2$ mass fraction is found to be orders of magnitude lower than for other molecules. Therefore, accounting for the SiO$_2$ condensation is negligible. 
 
As in BB20, we studied the solid formation close to the water ice line inside the water ice line (T>150 K) and outside the water ice line (T<150 K). The condensation temperatures for species involved by stoichiometric relations are the ones of \citet[][]{Lodders2003} (cf. also Table 1 from BB20).

\section{PBB composition as a function of metallicity}
\label{SecMassFractionMetallicity}

\begin{figure}
  \centering
    \includegraphics[width=8cm, clip=true, trim= 0cm 0cm 0cm 0cm]{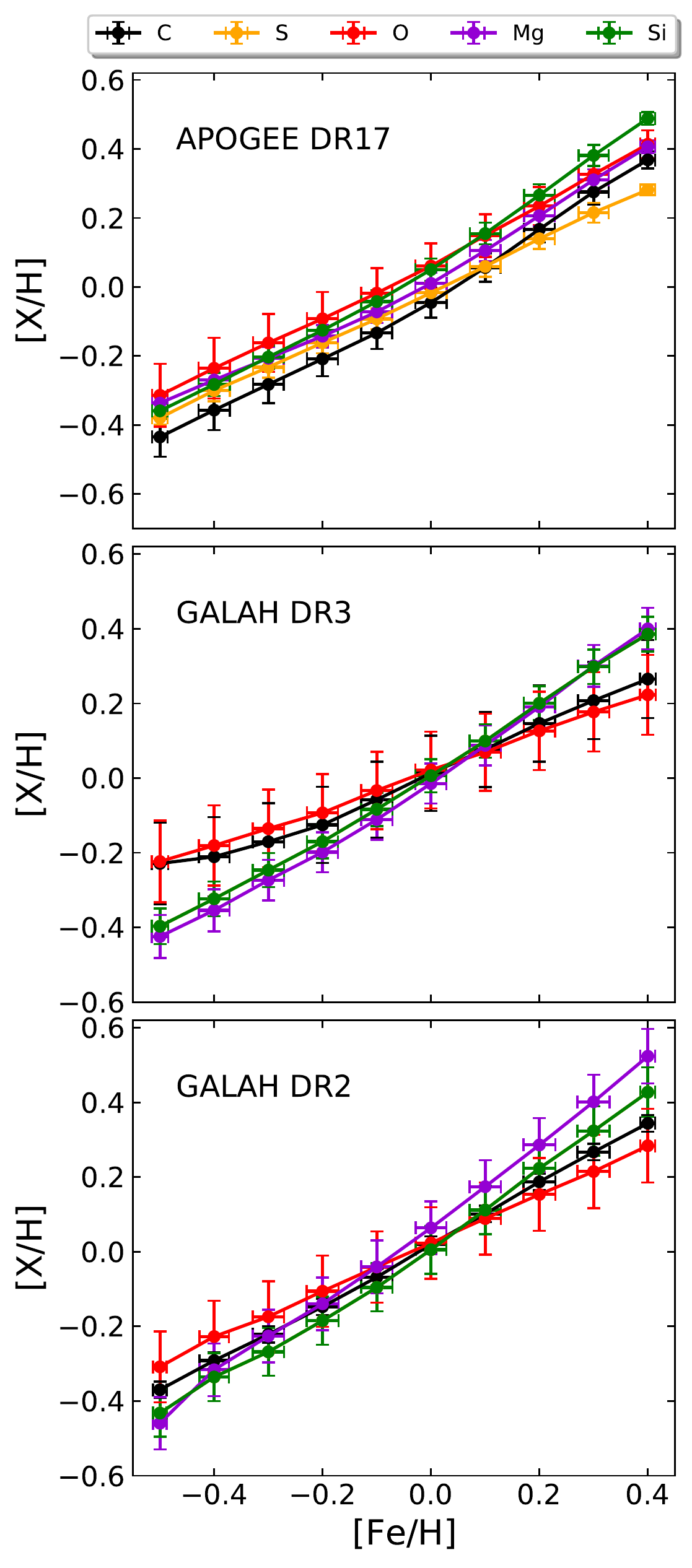}
     \caption{Stellar abundances of C, O, Mg, Si, and S as function of [Fe/H] in the observational surveys APOGEE-DR17, GALAH-DR3, and GALAH-DR2. The error bars are the mean deviations of the observations. In the GALAH samples, the sulphur scales in the same way as silicon.}
    \label{Fig0_XFe}
\end{figure}

In Fig. \ref{Fig0_XFe}, we plot the averaged stellar abundances [X/H] per bin of metallicity $\Delta$[Fe/H]=0.1 dex. We computed the average [X/H] per bin using only stars with good chemical element flags in Si, Mg, S, C, and O (we required each star to have all those quality flags). Because of weak statistics on extremely low and high metallicity bins, we limited our sample to -0.5<[Fe/H]<0.4 dex. We also required C/O<0.8 and Mg/Si<2. The number of selected stars for every survey is shown in Table \ref{Table2_SampleStars}.

\begin{table*}
\centering                                      
\begin{tabular}{cc|c|c|c|}          
                                                                                                              &                       & APOGEE-DR17      & GALAH-DR3             & GALAH-DR2         \\

\hline                                   

\multicolumn{1}{c|}{ }                                                                                   & Mg/Si<1        & 12 815 (20.5\%)       & 3646 (12.5\%)         &  179 (3.8\%)    \\
\multicolumn{1}{c|}{N$_{*}$ in Sect. \ref{SecMassFractionMetallicity}}   & 1<Mg/Si<2   & 49 796 (79.5\%)  & 25 719 (87.5\%)              &  4525 (96.2\%) \\

\hline                                   

\multicolumn{1}{c|}{ }                                                                                  &  Mg/Si<1       & 8518 (23.1\%)    & 3705 (12.4\%)               & /    \\
\multicolumn{1}{c|}{N$_{*}$ in Sect. \ref{SecDiag}}                                     &  1<Mg/Si<2   & 28 335 (76.9\%)  & 26 251 (87.6\%)               & /    \\

\hline                      

\end{tabular}

\caption{Sections \ref{SecMassFractionMetallicity} and \ref{SecDiag} require different selection criteria. Sect. \ref{SecMassFractionMetallicity} includes stars with -0.5 < [Fe/H] < 0.5 to average metallicity bins with significant statistics. For every star, good spectroscopic quality flags are required for Si, Mg, S (GALAH has no S abundances), C, and O. Sect. \ref{SecDiag} does not use metallicity cuts but requires good spectroscopic flags for every star for Si, Mg, S, C, and O, in addition to Ca and Ti, to compute [$\alpha$/Fe].}.

\label{Table2_SampleStars}
\end{table*}

\begin{figure*}
  \centering
    \includegraphics[width=6cm,clip=true, trim= 0cm 0cm 0cm 0cm]{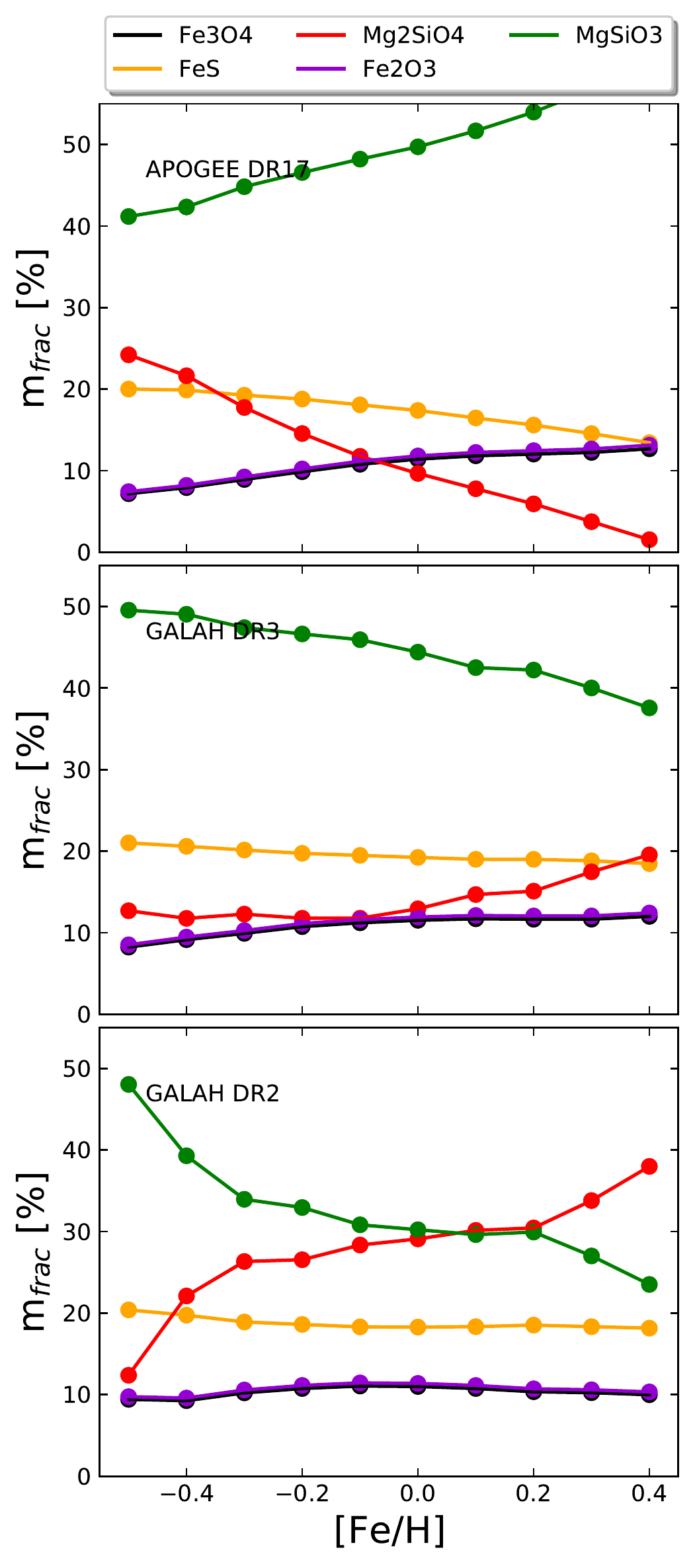}
    \includegraphics[width=6cm,clip=true, trim= 0cm 0cm 0cm 0cm]{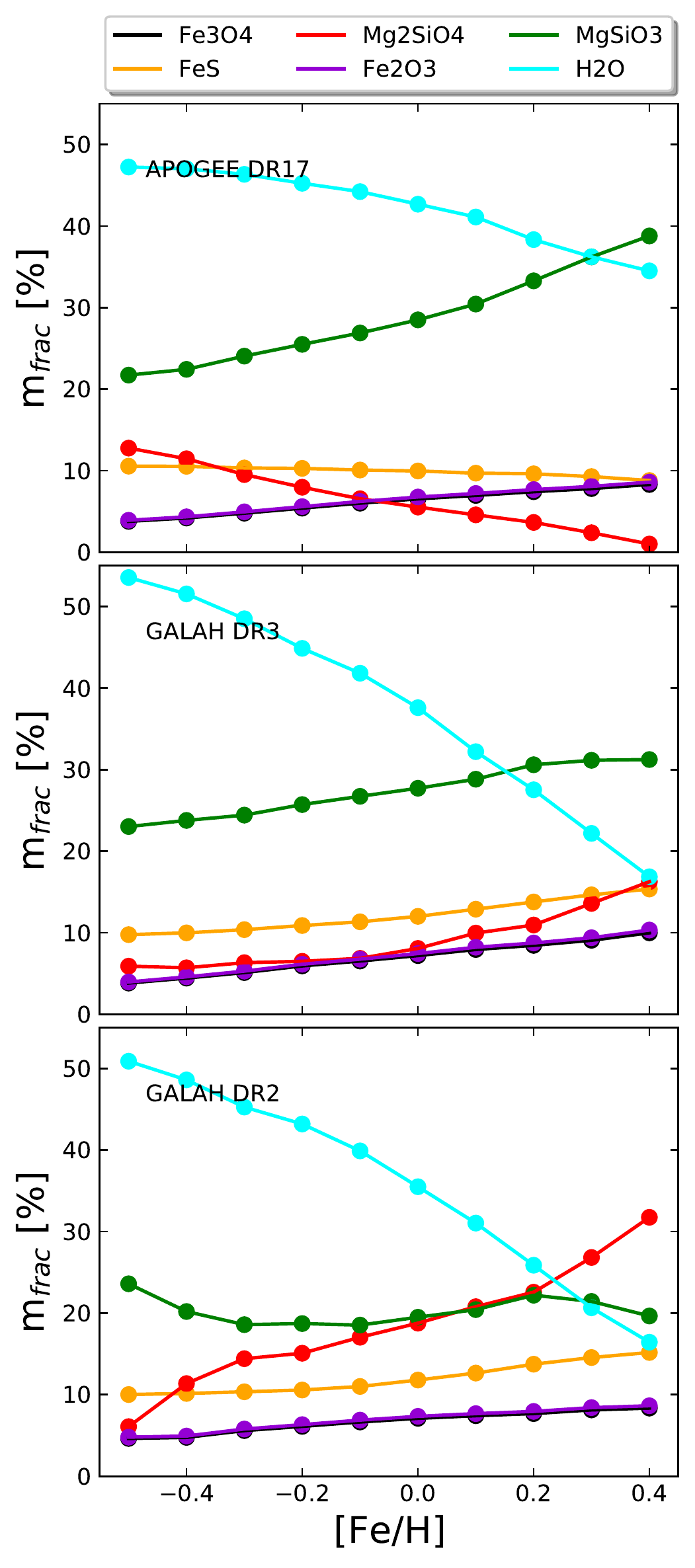}
    \caption{Mean molecular mass fractions per bin of metallicity. Top panels correspond to APOGEE-DR17; middle panels correspond to GALAH-DR3; bottom panels correspond to GALAH-DR2. Left panels correspond to the inner proto-planetary disc (T>150K); right panels correspond to the outer proto-planetary disc (T<150K, which includes H$_2$O ice). For the inner proto-planetary disc, the MgSiO$_3$ mass fraction is higher than the Mg$_2$SiO$_4$ mass fraction in APOGEE-DR17 and GALAH-DR3; albeit the trend with the metallicity is different. The MgSiO$_3$ and the Mg$_2$SiO$_4$ mass fractions are reversed between the GALAH-DR2 and GALAH-DR3 but only at [Fe/H] > 0. For [Fe/H]<0, the trend is similar. For the outer proto-planetary disc, the water ice mass fraction decreases with the metallicity in the three samples.}
    \label{Fig1_MfracFe}
\end{figure*}

Figure \ref{Fig1_MfracFe} shows the averaged PBB composition per bin of metallicity computed with the averaged stellar abundances [X/H] from Fig. \ref{Fig0_XFe}. The PBB mass fractions shown in Fig. \ref{Fig1_MfracFe} have been computed for the interior (T>150 K) and the exterior (T<150 K) of the water ice line.

The comparison of the APOGEE and the GALAH surveys show clear differences but also common trends in Fig. \ref{Fig1_MfracFe}. Both surveys show that inside the water ice line (T>150 K, left panels), the mass fractions of Fe-bearing molecules (FeS, Fe$_2$O$_3$ and Fe$_3$O$_4$) have very similar values, but with slightly different  behaviours between the APOGEE and the GALAH surveys. Indeed, FeS smoothly decreases, while Fe3O4 and Fe2O3 increase with [Fe/H] for the APOGEE survey; while no mass fraction dependency with [Fe/H]  is visible for the GALAH survey. The mass fraction values of Fe-bearing molecules are in the three surveys:  $\sim$20-15\% of FeS, $\sim$10\% Fe$_3$O$_4,$ and $\sim$10\% of Fe$_2$O$_3$.

For the Mg-bearing molecules the metallicity trend differs in the inner proto-planetary disc. The mass fractions of MgSiO$_3$ are increasing with [Fe/H] for APOGEE-DR17, but decreasing in both GALAH samples. The inverse trend is naturally found for Mg$_2$SiO$_4$. This can be explained, at first order, by the fact that for APOGEE-DR17, the averaged abundances give [Mg/H]<[Si/H] for large metallicities, while for GALAH-DR3 we obtain [Mg/H]$\sim$[Si/H] at large metallicities; for GALAH-DR2, Fig. \ref{Fig0_XFe} shows [Mg/H]>[Si/H]. This is an important difference between both surveys that impacts stoichiometric relations. However, we see that overall the Mg-bearing molecules dominate the PBB composition with approximately 50$\%$ of the total mass fraction at solar metallicity. This is common to the three surveys and appears to be a robust trend.

For T<150 K (right panel), the overabundance of oxygen enables the condensation of large amounts of water ice. The water ice is clearly dominating the PBB composition in metal-poor stars with [Fe/H]<0.1. We see a clear water ice-metallicity dependance; the water ice is decreasing with the metallicity in the three samples. As shown in Fig. \ref{Fig14_nCnO_AlphaFe}, the C/O ratio is increasing with [Fe/H] overall, which naturally explains the water ice-metallicity dependance. Moreover, for [Fe/H]>0, we have substantial differences of water content between APOGEE and GALAH. We discuss this point with more detail in Sect. \ref{SubSectWater}.

We also apply the stoichiometric relations to the GALAH-DR2 to compare with the work of BB20. After comparison with Fig. 10 of BB20, we see that the PBB mass fractions are similar and the overall trends of Fe-bearing molecules are consistent. However, we note that the cross between the MgSiO$_3$ and the Mg$_2$SiO$_4$ appears at different metallicity, respectively 0.1 in this study and -0.4  in BB20. These differences mainly come from the Mg/Si ratio that controls the Mg$_2$SiO$_4$/MgSiO$_3$ ratio. Both studies obtain [Mg/H]>[Si/H] at all metallicities, but the [Mg/H]-[Si/H] differences are larger in BB20, resulting in higher abundances of Mg$_2$SiO$_4$. In spite of these clear differences, the trends with [Fe/H] are comparable and the mass fraction values of Fe-bearing molecules are very similar.

The analysis is different when comparing the results obtained here with the updated GALAH-DR3 release and the ones obtained with GALAH-DR2 by BB20 and this work. For the inner proto-planetary disc (T>150 K) in the GALAH-DR2 sample, the MgSiO$_3$ mass fraction decreases with the metallicity (similarly to GALAH-DR3), but the mass fraction is lower than that of Mg$_2$SiO$_4$ for [Fe/H]<-0.1, which is not the case with GALAH-DR3. For the outer proto-planetary disc (T<150 K), the decreasing trend of water ice is very similar in DR2 and DR3. However, in DR2 the MgSiO$_3$ mass fraction remains lower than that of Mg$_2$SiO$_4$  for [Fe/H]<-0.1, while in DR3 the opposite is found for all [Fe/H]. 

In summary, we see a rather large modification of the order of $\sim$10\% in molecular mass fraction between the two last releases. This emphasises the need to analyse the spectroscopic data with utmost caution, even with the current high-quality surveys. Sect. \ref{SecErrors} aims to study the sensitivity of molecular mass fraction results in relation to the measured abundances uncertainties. Interestingly, for all surveys and data releases the water ice reduces as [Fe/H] increases.

\section{Distribution in the [$\alpha$/Fe]-[Fe/H] plane}
\label{SecDiag}

We aim to investigate the molecular mass fraction distribution in the [$\alpha$/Fe]-[Fe/H] diagram to explore the potential links between the galactic origin and the expected PBB composition. To compute the PBB composition distribution in the diagram [$\alpha$/Fe]-[Fe/H] diagram, we required every star to have good spectroscopic flags for Ca and Ti, in addition to the spectroscopic quality flags required in the previous sections: Mg, Si, S, C, and O. The only difference from Sect. \ref{SecMassFractionMetallicity} is the additional requirements on Ca and Ti to compute  [$\alpha$/Fe]. As in Sect. \ref{SecMassFractionMetallicity}, we require C/O<0.8 and Mg/Si<2. The number of selected stars is shown in Table \ref{Table2_SampleStars}, while Table \ref{Table1_GalComponents} summarises the classification in Galactic components for APOGEE and GALAH stars. 

\begin{table*}
\centering                                      
\begin{tabular}{ccc}          
\hline                                   
\hline                                   
                                                                & APOGEE-DR17                              & GALAH-DR3     \\
\hline                                   
\multicolumn{1}{c}{Thin}                                & 33457                                         &  26966                      \\
\multicolumn{1}{c}{Thick}                               & 1768                                          &  1521                        \\
\multicolumn{1}{c}{Intermediate}                & 1590                                          &  1468                       \\
\multicolumn{1}{c}{Halo}                                & 29                                                    &  1                                 \\
\hline                                   

\end{tabular}
\caption{Galactic population identification obtained from the kinematical classification applied to the sample used in Sect. \ref{SecDiag}.}
\label{Table1_GalComponents}
\end{table*}

\subsection{Mg and Si abundance distribution in the [$\alpha$/Fe]-[Fe/H] diagram}
\label{SecAbundancesAlphaFe}

\begin{figure*}
  \centering
                \includegraphics[width=5.9cm,clip=true,trim= 0cm 0cm 0cm 0cm]{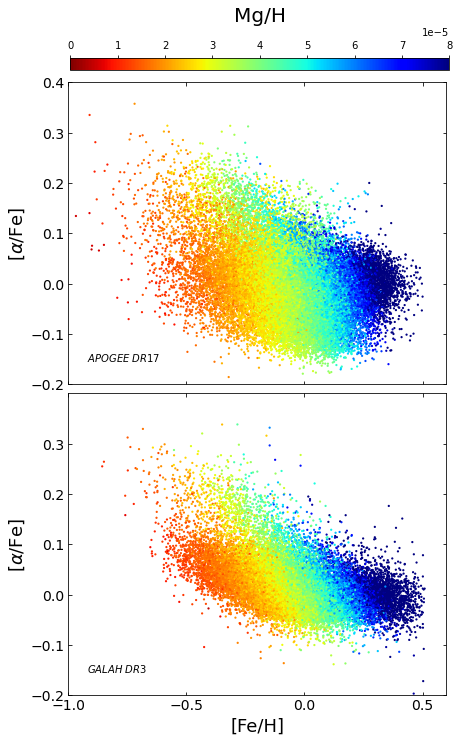}
                \includegraphics[width=5.9cm,clip=true,trim= 0cm 0cm 0cm 0cm]{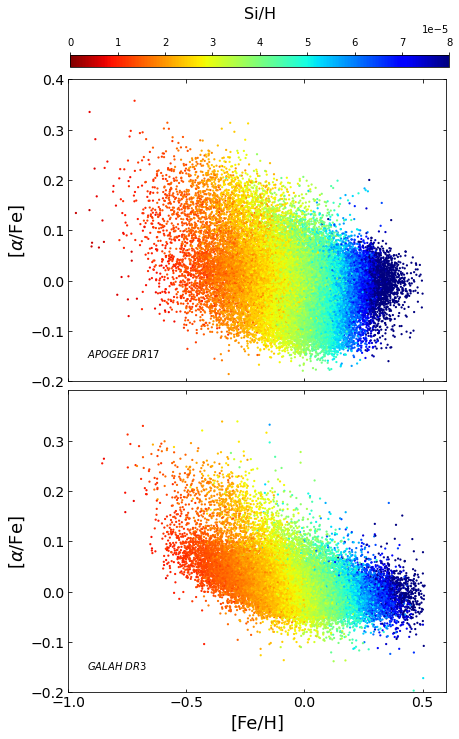}
                \includegraphics[width=6cm,clip=true,trim= 0cm 0cm 0cm 0cm]{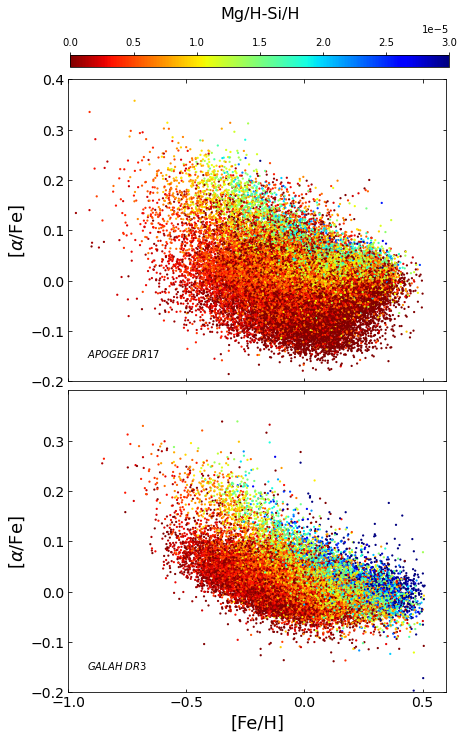}
                
     \caption{Stellar abundance distribution in the [$\alpha$/Fe]-[Fe/H] diagram for  Mg/H (left panel), Si/H (middle panel), and Mg/H-Si/H (right panel). We display the surveys APOGEE-DR17 (top panels) and GALAH-DR3 (bottom panels).}
     \label{Fig6_MgH_AlphaFe_FeH}
\end{figure*}

To understand the pattern distribution of the MgSiO$_3$ and Mg$_2$SiO$_4$ mass fractions in the [$\alpha$/Fe]-[Fe/H] diagram plane, we first show the chemical element abundances. Since the stoichiometric relations link the MgSiO$_3$ and Mg$_2$SiO$_4$ molecular abundances to the Mg and Si abundances, we compare, in Fig. \ref{Fig6_MgH_AlphaFe_FeH}, the pattern abundances of Mg/H (left panel), Si/H (middle panel), and Mg/H-Si/H (right panel) in the [$\alpha$/Fe]-[Fe/H] plane. Consistently with [Mg/H] and [Si/H] of Fig. \ref{Fig0_XFe}, the Mg/H and Si/H abundances increase with [Fe/H]\footnote{We remind the reader that A/B$\neq$[A/B]. X/H is the elemental ratio, while [X/H] is the solar-normalised logarithmic ratio.}. Figure \ref{Fig6_MgH_AlphaFe_FeH} shows that the Mg and the Si distributions follow a diagonal [$\alpha$/Fe]-[Fe/H] dependance. This is at some point expected since Mg and Si are included in the calculation of the alpha content [$\alpha$/Fe]. When [Mg/H] or [Si/H] increases, [$\alpha$/Fe] also increases. It appears that the general trends of Mg/H and Si/H are very similar for both surveys.

In the right panel of Fig. \ref{Fig6_MgH_AlphaFe_FeH}, we show the distribution of Mg/H-Si/H on the [$\alpha$/Fe]-[Fe/H] plane. The distribution still reveals a diagonal dependance, but it is quite different between both surveys. Despite the apparently weak differences between surveys for Mg/H and Si/H alone (left panels), this reveals something new. It is crucial since, as imposed by stoichiometric relations, the number of Mg$_2$SiO$_4$ molecules is directly linked to the Mg/H-Si/H difference. In other words, apparently subtle differences in chemical abundance patterns can have a significant impact on the PBB composition patterns.

\begin{figure*}
  \centering
                \includegraphics[width=8.2cm,clip=true,trim= 0cm 0cm 0cm 0cm]{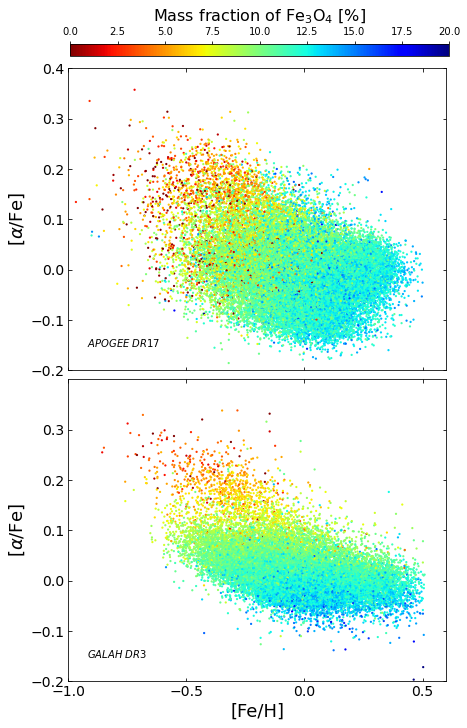}
                \includegraphics[width=8.1cm,clip=true,trim= 0cm 0cm 0cm 0cm]{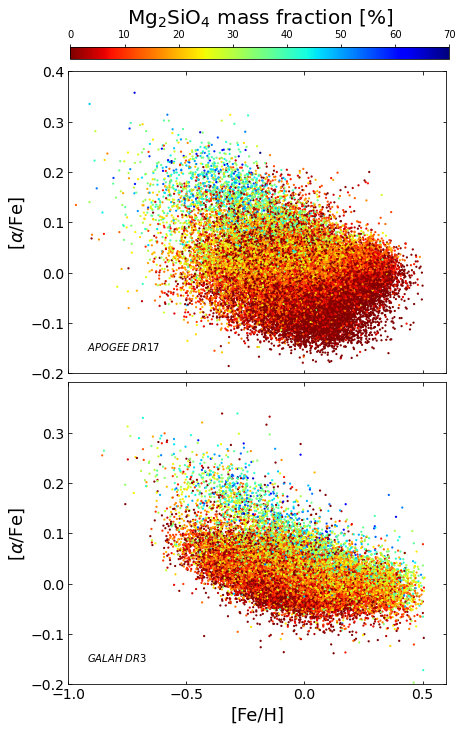}

     \caption{PBB mass fraction distribution of Fe$_3$O$_4$ (left panels) and Mg$_2$SiO$_3$ (right panels) for T>150 K (inner proto-planetary disc) in the [$\alpha$/Fe]-[Fe/H] diagram. We display the surveys APOGEE-DR17 (top panels) and GALAH-DR3 (bottom panels).}

    \label{Fig7_Mfrac_AlphaFe_Inner}
\end{figure*}

The pattern distribution shows a clear difference between the thin and the thick disc for the APOGEE stars, while the GALAH stars present high and low Mg/H-Si/H values, that is, rich and poor Mg$_2$SiO$_4$ abundances\footnote{Mg$_2$SiO$_4$ = Mg-Si is different than the Mg$_2$SiO$_4$ mass fraction.}, in both the thin and thick discs. The range of Mg/H-Si/H values is larger in GALAH (shown through the more extreme red and blue values in the right panel of Fig.4). Actually, the GALAH distribution has a stronger metallicity dependence. For [Fe/H]>0, the GALAH sample predicts higher values of Mg/H-Si/H than the APOGEE sample. Consequently, for [Fe/H]>0, we can expect the GALAH sample to have PBBs with higher Mg$_2$SiO$_4$ mass fractions than the APOGEE sample (see right panel of Fig. \ref{Fig7_Mfrac_AlphaFe_Inner}).

Despite local differences, the overall trends are still similar. In particular, the alpha-rich stars usually associated with the thin disc are predicted to be richer in Mg$_2$SiO$_4$ than the thick disc. It is important to note that the Mg/H and Si/H chemical distribution (left panels of Fig. \ref{Fig6_MgH_AlphaFe_FeH}) have an [$\alpha$/Fe] dependance weaker than the [Fe/H] dependance. However, the final PBB composition (right panel of Fig. \ref{Fig6_MgH_AlphaFe_FeH}) dependance is with both the $\alpha$-content [$\alpha$/Fe] and the iron content [Fe/H]. Consequently, we obtain clear differences between the thin and thick disc stars.

\begin{figure}
  \centering
                \includegraphics[width=8cm,clip=true,trim= 0cm 0cm 0cm 0cm]{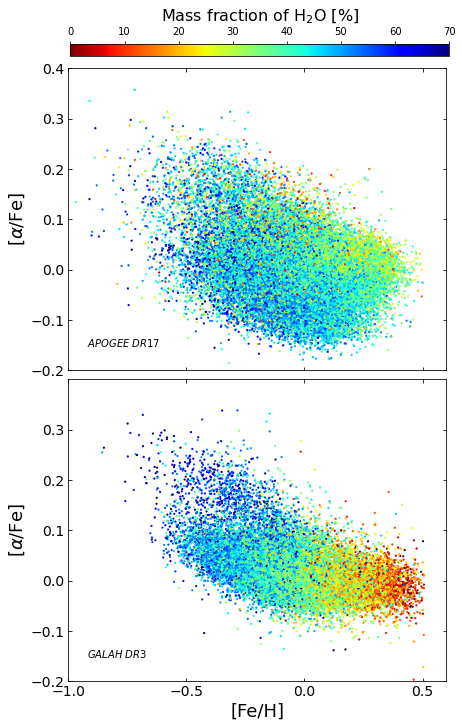}

     \caption{PBB mass fraction distribution of H$_2$O for T<150 K (outer proto-planetary disc) in the [$\alpha$/Fe]-[Fe/H] diagram. We display the surveys APOGEE-DR17 (top panels) and GALAH-DR3 (bottom panels).}

    \label{Fig7_Mfrac_AlphaFe_Outer}
\end{figure}

\subsection{Mass fraction distribution in the [$\alpha$/Fe]-[Fe/H] plane}
\label{SecMassDiagAlphaFe}

\subsubsection{Inner proto-planetary disc (T>150 K):  Fe$_3$O$_4$ and Mg$_2$SiO$_4$} 

As illustrative examples, we plot mass fraction distribution of Fe$_3$O$_4$ and Mg$_2$SiO$_4,$ respectively, in the left and right panels of Fig. \ref{Fig7_Mfrac_AlphaFe_Inner}. The Fe$_3$O$_4$ mass fraction ranges from 0 to $\sim$20\%, while the Mg$_2$SiO$_4$ mass fraction ranges from 0 to $\sim$70\%. For both surveys, the alpha-rich stars have higher Mg$_2$SiO$_4$ and lower Fe$_3$O$_4$ mass fractions than alpha-poor stars. The trend of a higher Mg$_2$SiO$_4$ and lower Fe$_3$O$_4$ means the mass fraction for the thick disc is also seen in the histograms in Appendix  \ref{FigAppendixHistoKine}. For Mg$_2$SiO$_4$, Fig. \ref{FigAppendixHistoKine}  and table \ref{Table3_MeanMfracCompGal} show that the difference between the thin and thick discs is of the order of 9\%, and for Fe$_3$O$_4$ we have a difference of 4\%.

We note that in Fig. \ref{Fig1_MfracFe}  the Fe$_3$O$_4$ mass fraction per bin of [Fe/H] is almost constant for the GALAH samples. However, in Fig. \ref{Fig7_Mfrac_AlphaFe_Inner} we now see a clear dependance with the $\alpha$-content in the [$\alpha$/Fe]-[Fe/H] diagram. In other words, the molecular mass fraction per bin [Fe/H] hides the PBB composition we may expect for a given star. As quantified in Fig. \ref{Fig_Mfrac_ThinThickperFeH}, the mass fraction in the thin and the thick discs may be quite different, in particular for Mg-bearing molecules. The trend with metallicity is, however, similar between both the thin and the thick discs, at the inner proto-planetary disc (T>150 K, left panels) and the outer proto-planetary disc (T<150 K, right panels). We note that using the chemical classification method (cf. Appendix \ref{AppendixDensityPlot}), the mass fraction values are different but the comparisons between the thin and thick discs' PBBs are very similar. As shown by Fig. \ref{FigMfracFeH_ChemicalClass}, the Mg$_2$SiO$_4$ mass fraction is higher in the thick disc than in the thin disc, consistently with Fig. \ref{Fig_Mfrac_ThinThickperFeH}. Actually, for all molecules the overall trends are equivalent in both classification methods.

Consistently with the right panel of Fig  \ref{Fig6_MgH_AlphaFe_FeH}, both surveys have very different Mg$_2$SiO$_4$ mass fraction values for [Fe/H]>0 in Fig. \ref{Fig7_Mfrac_AlphaFe_Inner}. Overall, the molecular pattern distribution appears to be slightly different, but the general trends are similar for both surveys. We found that the molecular mass fraction has a strong dependance with [$\alpha$/Fe], suggesting that the different galactic populations should have different PBB compositions. It becomes clear that the mass fraction calculations averaged per bin of [Fe/H] may hide important trend information. Using the average values might be dangerous to draw some direct conclusions. Stellar abundances have to be analysed individually.

\subsubsection{Outer proto-planetary disc (T<150 K): H$_2$O} 
\label{SubSectWater}

\begin{figure*}
 \centering
        \includegraphics[width=8.3cm,clip=true,trim= 0cm 0cm 0cm 0cm]{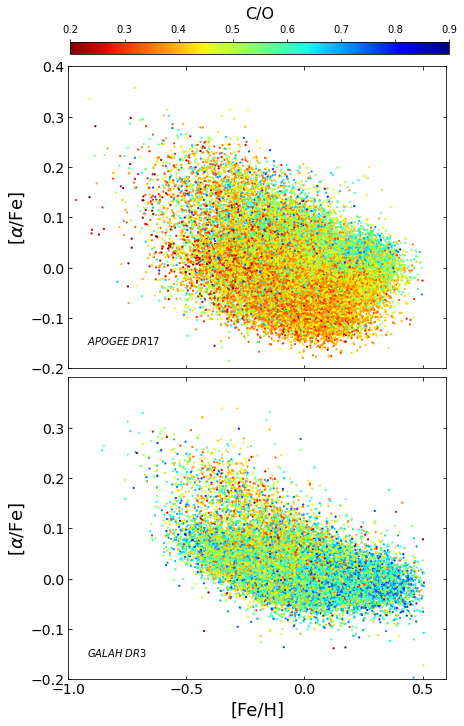}
        \includegraphics[width=8cm,clip=true,trim= 0cm 0cm 0cm 0cm]{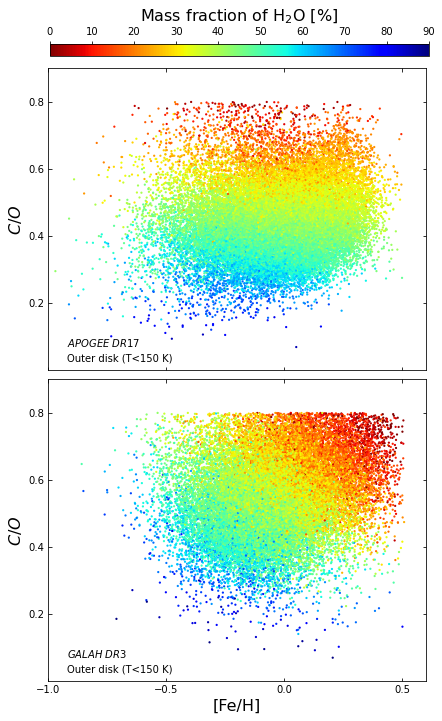}
     \caption{C/O distribution in the [$\alpha$/Fe]-[Fe/H] diagram (left panel) and the mass fraction of H$_2$O as a function of C/O and [Fe/H]. We display the surveys APOGEE-DR17 (top panels) and GALAH-DR3 (bottom panels).}    
    \label{Fig14_nCnO_AlphaFe}
\end{figure*}

Figure \ref{Fig7_Mfrac_AlphaFe_Outer} shows the H$_2$O mass fraction distribution in the [$\alpha$/Fe]-[Fe/H] plane. The GALAH sample has a clear metallicity dependence while for APOGEE this dependence is weaker. In both samples there is also an alpha dependance but to a lesser degree. The water mass fraction distribution is thought to be linked to the C/O. The lower the C/O ratio, the more oxygen is available to be condensed as water molecules. However, we check here whether the C/O distribution matches with the H$_2$O mass fraction distribution. In the case of APOGEE, we see that the water pattern distribution in Fig. \ref{Fig7_Mfrac_AlphaFe_Outer} matches with the C/O pattern distribution in the left panel of Fig. \ref{Fig14_nCnO_AlphaFe}. This is consistent with the right panel of Fig. \ref{Fig14_nCnO_AlphaFe} where the C/O and the water mass fraction are perfectly correlated. This is not the case for GALAH, where Fig. \ref{Fig7_Mfrac_AlphaFe_Outer} shows a clear anti-correlation of water with metallicity, which is less obvious for C/O in the [$\alpha$/Fe]-[Fe/H] plane (left panel of Fig. \ref{Fig14_nCnO_AlphaFe}). This is because the water mass fraction is not perfectly correlated with C/O (right panel of Fig. \ref{Fig14_nCnO_AlphaFe}). For GALAH, the water mass fraction is more complex than a simple correlation with C/O. There is a metallicity dependence that is not observed in APOGEE. 

Both surveys show a common trend: the water mass fraction decreases with metallicity. This is shown in Fig. \ref{Fig_Mfrac_ThinThickperFeH} for both stellar populations: thin and thick disc (classified by the kinematical method). This is also confirmed using the chemical classification method in Appendix \ref{AppendixDensityPlot} and Fig. \ref{FigMfracFeH_ChemicalClass}. It can be explained by the overall increases of C/O with metallicity (right panel of Fig.\ref{Fig14_nCnO_AlphaFe}), such that the water mass fraction decreases with metallicity. For GALAH, the anti-correlation between the water content and the metallicity is strong because H$_2$O is correlated with C/O but also with [Fe/H]. For APOGEE, the anti-correlation between H$_2$O and [Fe/H] is weaker than for GALAH because the water content is only correlated with C/O.

The anti-correlation between water and metallicity is common to both surveys and may suggest that metal poor stars are highly likely to have water-rich PBBs. However, the separation of the thin and thick discs is not so clear in terms of water mass fraction. Fig. \ref{Fig_Mfrac_ThinThickperFeH} suggests that for APOGEE the thin disc is more suitable for water-rich PBBs, while for GALAH it is the thick disc that should host water-rich PBBs. These results show the importance of obtaining spectroscopic measurement with small errors for oxygen and carbon.

\subsection{Bimodal PBB composition}
\label{wivalley}

By analysing 371 HARPS stars, S17 found a difference between the PBB composition of the thin disc, thick disc and high-alpha, metal-rich, and halo stars. With a more robust statistics, and taking advantage of stellar synthetic population models, C19 studied the PBB composition in the [$\alpha$/Fe]-[Fe/H] diagram. They highlight the correlation of the PBB composition with the alpha content for typical values of [$\alpha$/Fe] and [Fe/H]. They found that the thick disc, thin disc, the halo, and the bulge stars have different PBB compositions. With the HYPATIA\footnote{https://www.hypatiacatalog.com/} catalogue, \cite{Michel2020} found similar results to S17 and C19 for the thin and the thick disc PBB compositions.

In particular, C19 found a synthetic water/iron gap in histograms of their Fig. 2 (for galactic distances up to 100 pc) and their Fig. 3 (for galactic distance up to 50 kpc). This bimodal PBB composition appears because the iron and the water mass fractions correlate with [$\alpha$/Fe] (as seen in Figs. \ref{Fig7_Mfrac_AlphaFe_Inner} and \ref{Fig7_Mfrac_AlphaFe_Outer} ). Then, the known dip density of stars in the region between the alpha-rich thick disc and the alpha-poor thin disc generates a dip of stars with intermediate PBB compositions.

\begin{figure*}
  \centering
                \includegraphics[width=9cm,clip=true,trim= 0cm 0cm 0cm 0cm]{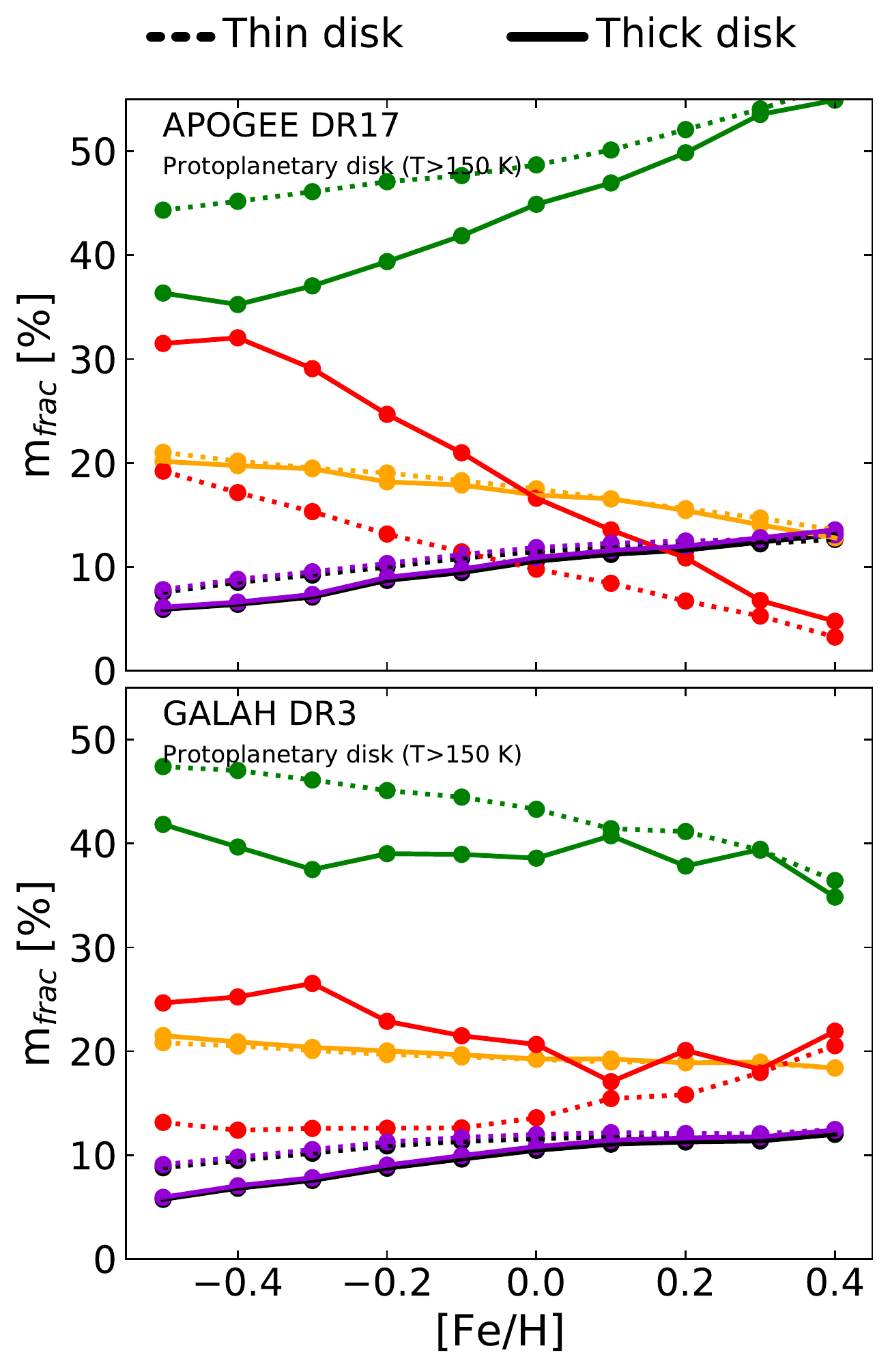}
                \includegraphics[width=9cm,clip=true,trim= 0cm 0cm 0cm 0cm]{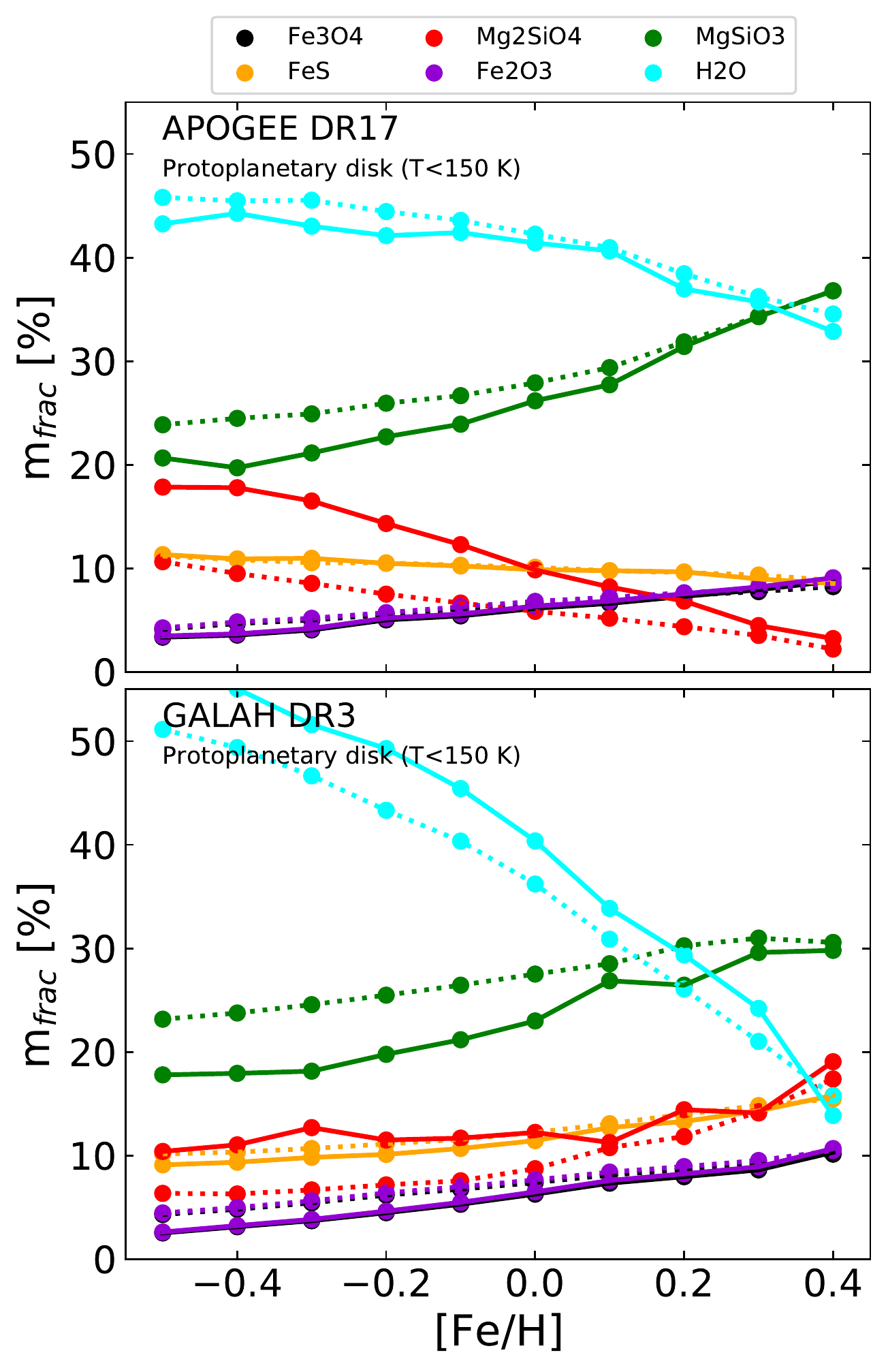}

     \caption{Mean molecular mass fractions per bin of metallicity for thin disc (solid line) and thick disc (dashed line). Top panels correspond to APOGEE-DR17; bottom panels correspond to GALAH-DR3. Left panels correspond to the inner proto-planetary disc (T>150K); right panels correspond to the outer proto-planetary disc (T<150K, which includes the H2O molecule).}

    \label{Fig_Mfrac_ThinThickperFeH}
\end{figure*}

The results of the present study validate most of the general trends found in C19.  In particular, Fe-bearing and Mg-bearing molecule mass fractions are found to correlate with the stellar alpha content. The two large survey analyses tend to show that the PBB composition differences between the galactic populations are robust. The case of water ice is more difficult to discuss and cannot be directly compared with C19. First, C19 computed a synthetic stellar population including a large number of stars in the thin and the thick discs, while the present sample has a comparatively lower thick/thin star ratio. Moreover, the stoichiometric model used in C19 does not consider that the carbon molecules can bind oxygen. In addition, the water pattern distribution computed here is not so clear when comparing both samples, APOGEE and GALAH, in Fig. \ref{Fig7_Mfrac_AlphaFe_Outer}. The water mass fraction dependance with alpha-content is uncertain. This can be seen from Fig. \ref{Fig_Mfrac_ThinThickperFeH}, where the GALAH thin disc has a lower PBB water content than the thick disc for the all metallicity bins, while the opposite is found for the APOGEE sample. This is also illustrated by the histograms in Appendix \ref{FigAppendixHistoKine}, where the mean water mass fraction of the thick disc is higher for the GALAH DR3 (consistently with C19) but lower for APOGEE DR17. Finally, using the chemical classification method in Appendix \ref{AppendixDensityPlot} with Fig. \ref{FigMfracFeH_ChemicalClass}, we also find similar results.

\section{Propagation error study}
\label{SecErrors}

The PBB composition is directly determined from the spectroscopic abundances. However, even with relatively high-resolution spectroscopic data, the measured uncertainties can be large. This implies large differences in PBB composition, such as the ones obtained among GALAH-DR2, GALAH-DR3, and APOGEE-DR17. These results motivate the following propagation study error. We aim to estimate how robust our conclusions on PBB composition could be; in other words, how could the molecular mass fraction trends be modified when taking into account the error bars in spectroscopic abundances? 

For simplicity, we only used the GALAH-DR3 survey because the typical error bars are similar to those of APOGEE. Moreover, because we aim to discuss orders of magnitude, we only focus on the PBB composition per bin of iron content as in Sect. \ref{SecMassFractionMetallicity}. For similar reasons, we exclude the very low proportion of stars with Mg/Si<1 (excluding SiO$_2$ molecules) and Mg/Si>2. As an illustrative example, we discuss the observed upper and lower limits of [Mg/Fe] and [O/Fe]. Appendix \ref{AppendixErr} shows additional results obtained for the other chemical elements.  

\begin{figure}
  \centering
    \includegraphics[width=8cm,clip=true, trim= 0cm 0cm 0cm 0cm]{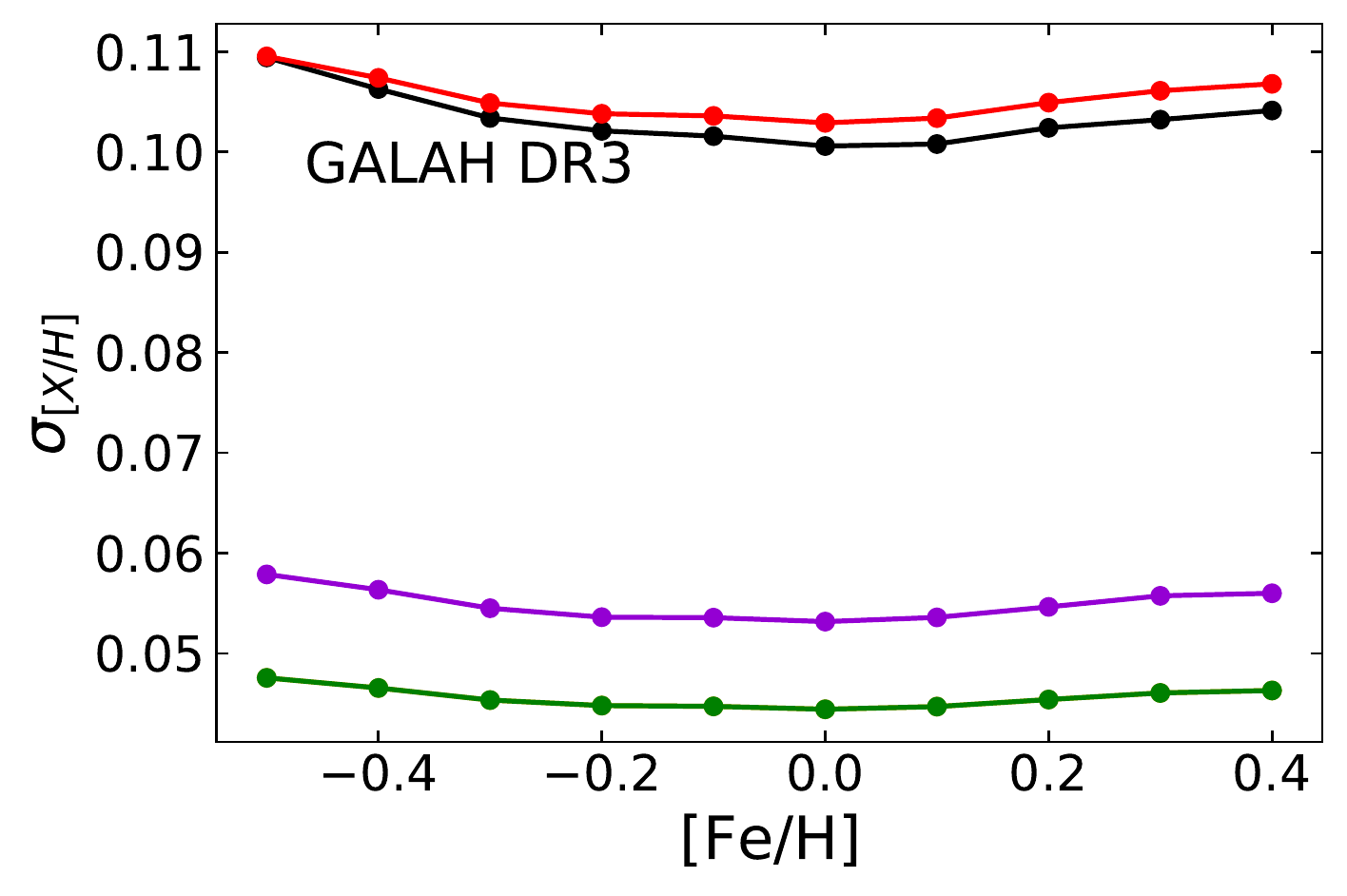}
     \caption{Averaged uncertainties per bin of [Fe/H] for GALAH-DR3. Colour-code is the same as in Fig. \ref{Fig0_XFe}. Clearly, the error bars for carbon and oxygen are much larger compared to those of Mg and Si.}
    \label{Fig3_ErrXH}
\end{figure}

From the stoichiometric model (Sect. \ref{SecChemicalModel}), we know that the mass fraction trends of Mg-bearing molecules are related to [Mg/H] and [Si/H] abundances. So, the propagation error study basically varies [Mg/H] and [Si/H] to evaluate the impact on the resulting molecular mass fraction. We varied [X/H] using the mean error bars from GALAH-DR3, such as \([X/H]_{test} = [X/H]+ \sigma_{[X/H]}\). The observed averaged uncertainties \(\sigma_{[X/H]}\) per bin of iron metallicity are plotted in Fig. \ref{Fig3_ErrXH}. We see indeed that oxygen and the carbon abundances have large spectroscopic error measures. Moreover, the dependance with [Fe/H] is weak in this selected sample. As clarified by a private communication, the error bars noted e$\_$x$\_$fe in the GALAH-DR3 surveys actually correspond to the error bars on [X/H] that we note here: \(\sigma_{[X/H]}\).

\subsection{Inner proto-planetary disc (T>150 K)}

\begin{figure*}
  \centering
    \includegraphics[width=8cm,clip=true, trim= 0cm 0cm 0cm 0cm]{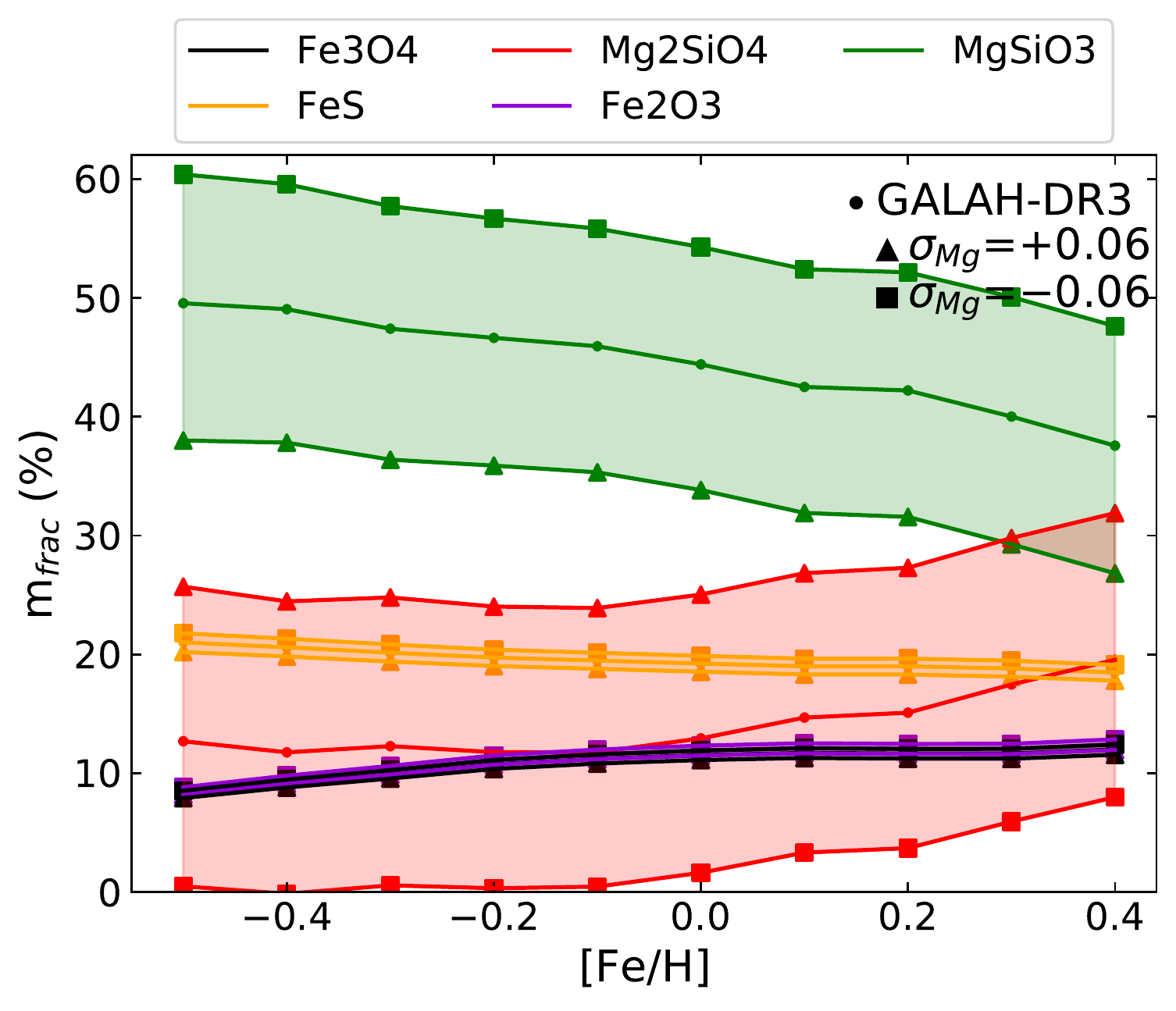}
    \includegraphics[width=8cm,clip=true, trim= 0cm 0cm 0cm 0cm]{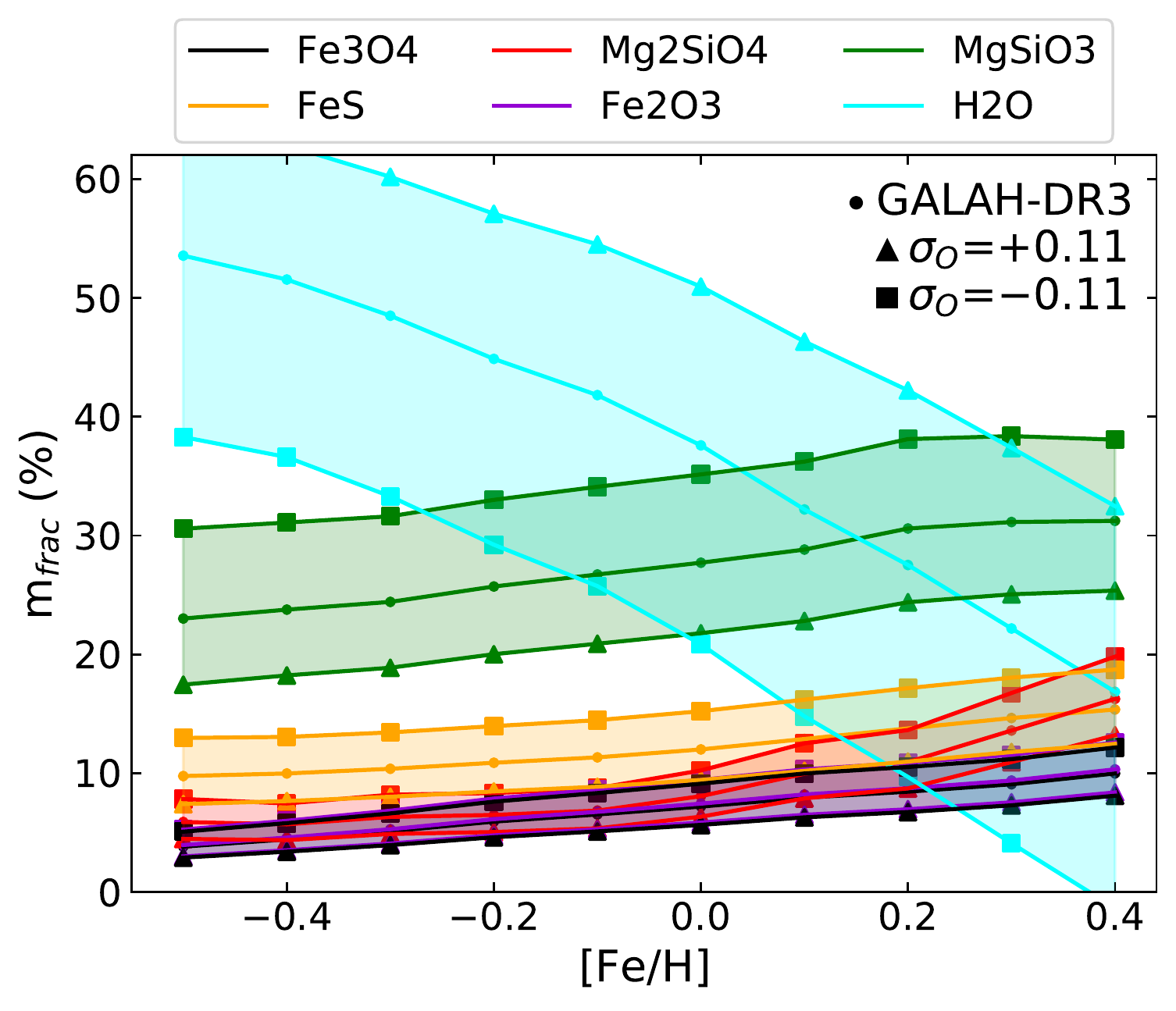}
     \caption{Mean PBB mass fractions per bin of [Fe/H] for GALAH-DR3 for modified abundances. Left panel considers the inner proto-planetary disc (T>150 K) varying \( [Mg/H] _{test} = [Mg/H] \pm \sigma_{Mg}  \). Right panel considers the outer proto-planetary disc (T<150 K) varying \( [O/H] _{test} = [O/H] \pm \sigma_{O}  \).}
    \label{2_ErrMfrac_Fe}
\end{figure*}

The left panel of Fig. \ref{2_ErrMfrac_Fe} illustrates the impact of stellar upper limit abundances on mass fractions in the inner proto-planetary disc when varying \([Mg/H]_{test}\) abundances such as  \([Mg/H]_{test} = [Mg/H]+ \sigma_{[Mg/H]}\). For this test, we used \(\sigma_{[Mg/H]} =\pm0.06,\) as observed in Fig. \ref{Fig3_ErrXH}. As expected when decreasing \([Mg/H]_{test}\) by \(\sigma_{[Mg/H]}$$=$$-0.06\), the MgSiO$_3$ mass fraction increases, while the Mg$_2$SiO$_4$ mass fraction decreases. This is an increase (decrease) of $\sim$10\% of the total mass fraction. Even in the upper error bar of \(\sigma_{[Mg/H]}$$=$$+0.06\) (triangle points), the Mg$_2$SiO$_4$ mass fraction is predicted to remain in lower proportions than MgSiO$_3$. We may expect a differential mass fraction between Mg$_2$SiO$_4$ and MgSiO$_3$ of 60\% for metal-poor stars and 40\% for metal-rich stars. The proportion of solids containing Fe is not modified because we only modified the Mg abundances in this plot. 

In the left panels of Fig. \ref{FigAppendix}  (cf. Appendix \ref{AppendixErr}), we show the propagation error varying [Si/H]. When decreasing \([Si/H]_{test,}\) the MgSiO$_3$ mass fraction decreases, while the Mg$_2$SiO$_4$ mass fraction increases. We stress that in case of varying Si and Mg abundances in opposite sign, the impact on PBB composition would be similar.

\subsection{Outer proto-planetary disc (T<150 K)}

In the right panel of Fig. \ref{2_ErrMfrac_Fe}, we focus on the outer proto-planetary  disc. We compute the propagation uncertainties on the molecular mass fraction varying the oxygen abundances [O/H]. We plotted all molecules condensed in the outer proto-planetary disc including H$_2$O. Oxygen is varied by \(\sigma_{[O/H]}$$=$$\pm0.11,\) as shown in Fig. \ref{Fig3_ErrXH}. A higher abundance of oxygen will naturally facilitate water ice condensation. The water ice mass fraction is increased (decreased) by 15\%, hence a difference of $\sim$30\% mass fraction between limit cases.

\subsection{Summary}

We did a propagation error test using the GALAH sample to determine the error we could expect on PBB composition predictions. This simple test shows that the exact mass fraction values should be discussed with caution because of the large uncertainties. These results of course emphasise the limitations of actual spectroscopic observations when they are used to predict the PBB composition in stellar populations. However, we believe that the overall trends are robust enough to be discussed.

\section{How water impacts planet populations across the Galaxy}
\label{SecRadius}

Planet formation models tend to show that the planet and the host star compositions are correlated \citep[e.g.][]{Bitsch2020}. In this line, combining observational data and internal planet structure models, \cite{Adibekyan2021} found that the stellar composition is not one-to-one with the planet composition, but it is highly correlated. Assuming that the chemical host star composition imprints the population of planets in the Galaxy, the question remains as to how different those planets around the thick and the thin disc host stars can be.

From the results obtained here with GALAH and APOGEE, we can reasonably expect that water-rich planets are more likely found around a metal-poor host. Indeed, as shown above, the chemical abundances in GALAH-DR3 and APOGEE-DR17 imply an anti-correlation between metallicity and water abundance (see Sect. \ref{SubSectWater}, Fig. \ref{Fig7_Mfrac_AlphaFe_Outer} and Fig. \ref{Fig_Mfrac_ThinThickperFeH}). However, whether the water abundance of PBB is a good proxy for planets or whether the thick disc stars are suitable hosts of water worlds and/or hycean planets remains a matter of debate.

On one hand, observations show that the occurrence rates of gas giants increase with the host star metallicity \citep[e.g.][]{Gonzalez1997,Santos2004,FischerValenti2005,Sousa2008,Johnson2010}. These observations support planet formation simulations showing that the metallicity increases the gas giant formation efficiency \citep[e.g.][]{IdaLin2004,Mordasini2012,Ndugu2018}. On the other hand, icy pebbles are thought to enable faster planet formation. In the case of the Solar System, \cite{Morbidelli2015} proposed that the dichotomy between rocky and giant planets results from the original dichotomy between dry and icy pebbles. Because of the water sublimation, they assume that the pebble flux is different by a factor of $\sim$2 inside and outside the ice line. The smaller grains (millimetre-sized) inside the ice line result in very different growth rates of the terrestrial planets and the giants.

Both parameters -metallicity and water mass fraction- are anti-correlated from stoichiometric relations. An interesting question to ask is if the water content could compensate a decrease of metallicity in the same way the alpha elements seems to do for low metallic planet host stars \citep{BashiZucker2019, Bashi2020}. A priori, a minimal metallicity seems necessary to form planets \citep[e.g.][]{JohnsonLi2012}, but the role of water in planet formation is an open topic \citep[e.g.][]{MullerSavvidouBitsch2021}. We briefly discuss this point in Sect. \ref{SecAlphaWaterImpact}.

\subsection{Are water worlds more likely around thin or thick disc host stars?}

Recently, \cite{Ghezzi2021} found that planets larger than Neptune, R$_p$>=4.4 R$_\oplus$, are only observed for super-solar metallicities. This is in line with previous observational \citep[e.g.][]{Gonzalez1997,Santos2004,FischerValenti2005,Sousa2008,Johnson2010} and theoretical  \citep[e.g.][]{IdaLin2004,Mordasini2012,Ndugu2018} results that tend to show that giant planets form easier in metal-rich stars. Instead, for sub-solar metallicities stars, \cite{Ghezzi2021} observed only sub-Neptune planets R$_p$<=4.4 R$_\oplus$. Although giant planets have been observed around thick disc host stars \citep{Haywood2009,DelgadoMena2021}, the results of \cite{Ghezzi2021} may indicate that the majority of planets orbiting thick disc hosts are sub-Neptunes (rocky or planets with small gaseous envelopes) since thick disc stars are rather metal-poor.

Assuming a clear correlation between PBBs and planet composition, we may expect a non-negligible proportion of sub-Neptunes to be water-rich (since those planets are small planets orbiting metal-poor stars). However, it is not clear which of the thin or thick disc host stars may be more suitable for water worlds and/or the hycean worlds also predicted by other works \citep[][]{Madhusudhan2021}. As seen in the right panel of Fig. \ref{Fig_Mfrac_ThinThickperFeH}, the PBB water content has a clear trend with metallicity, but the thin/thick disc differences are not obvious. Finally, the extrapolation for planets is based on the assumption that the water content of initial PBB correlates with the water content of final planet for low metallicity host stars. This a rather simplified point of view that needs to be proven from observations and theory.

\subsection{Does the water play a role similar to the alpha elements for low-metallicity planet host stars?}
\label{SecAlphaWaterImpact}

Interestingly, it has been observed that most of the metal-poor planet host stars, [Fe/H]<-0.3, belong to the thick disc population \citep{Haywood2008,Haywood2009}. The number of statistics is still small, but the trend exists for ten Neptune-like planets and around five Jupiter-like planets \citep{Adibekyan2012a}. As discussed by \cite{Adibekyan2012a} \citep[see also][]{Gonzalez2009}, the planet incidence may be linked to refractory elements (as Mg, Si, and Fe) and not necessarily to iron only. The idea is that the overabundance of refractory elements of the thick disc stars may compensate their weak iron abundance in order to form planets \citep{BashiZucker2019, Bashi2020}. Moreover, \cite{Chen2022} found that the largest planets (sub-Neptunes) in the planet radius valley \citep{Fulton2017} are preferentially around $\alpha$-rich host stars. Despite this $\alpha$ dependence, they did not find a clear trend with the galactic origin. That is to say that the sub-Neptunes are found around $\alpha$-rich stars indistinctly from the thin and the thick discs.

The predicted H$_2$O mass fraction is higher for thick stars compared to the thin disc stars in GALAH DR3 (Fig. \ref{Fig_Mfrac_ThinThickperFeH}), consistently with the trend observed in S17 with the HARPS-GTO sample. As a consequence, we may speculate that for very low metallicities the water content plays a role. In particular, we suggest that planet formation models should investigate whether the thick disc host stars form planets more easily than thin disc host, due to the presence not only of refractory elements but also because of water-rich PBs. An environment with an overabundance of both water and refractory elements may be a potential explanation to the greater planet incidence among the thick disc population than among the thin disc one for [Fe/H] <-0.3 dex.

Indeed, planet formation models show that water can help to form planets more easily because icy pebbles are thought to be accreted faster than dry pebbles \citep{Morbidelli2015}, because water ice pebbles can stick better \citep{GundlachBlum2015}, and because water can condense and facilitate the growth of pebbles \citep{RosJohansen2013, RosJohansen2019}. However, we need to keep in mind that the larger icy pebbles cannot compensate for lower metallicity in the accretion efficiency. In pebble accretion, the Stokes number $S_t$ (measurement for the particle size) scales with the factor of 2/3 on the accretion rate, while the pebble surface density scales with a factor of 1 on the accretion rate \citep[e.g.][]{Lambrechts2014, Bitsch2015}. However, for similarly low metallicities, [Fe/H] <-0.3 dex, the alpha-rich and water-rich PBB host stars may have an advantage to form planets. It is thus clear that the detailed chemical compositions of stars is needed to understand when and where planet formation started in our galaxy.

\subsection{Composition of interstellar objects}

One other expected impact is on the composition of small bodies formed together with planets, but not completely consumed by the planet formation process. Our results clearly indicate that the host star abundance can have significant influences on the water content of small bodies found in their planetary systems, with water-rich small bodies preferentially formed around metal-poor stars, and water-poor small bodies formed around metal-rich stars. Several mechanisms can allow for these small bodies to be stripped from their home systems. Similarly to our own system, the bulk of small bodies would be ejected during the early phases of the system’s formation \citep[e.g.][]{Raymond2020}. Our results thus support the two populations of interstellar objects found by \cite{Lintott2022}. 

If we can demonstrate that the properties of interstellar objects observed in the Solar System are representative of their characteristics inherited from their home systems (i.e. that they are not too significantly altered before they reach the Solar System), these objects will provide a different line of investigation for constraining the star and planet formation processes. For that, those interstellar objects should not be altered by their pathway across the Galaxy and the physical constrain at their own system \citep{Guilbert-Lepoutre2011,Guilbert-Lepoutre2012,Guilbert-Lepoutre2014}.

\section{Conclusion}
\label{Conclu}

The stoichiometric models allow us to reveal the PBB composition. They assume a condensation phase in a homogenous gaseous disc composition approximated by the atmospheric stellar abundances. This approach enables to discuss the PBB composition trends in the stellar population of the Galaxy. 

Our goal here is to predict the PBB compositions from large spectroscopic surveys to fully exploit the incredible amount of observational data. We decided to analyse the APOGEE and GALAH surveys separately because the determination of their stellar compositions are not necessarily derived in an homogeneous way. This avoids mixing data from different surveys, which may include different intrinsic biases that are hard to disentangle. In addition, the choice of large surveys decreases the potential biases due to a high number of statistics.

We combined the APOGEE-DR17 and GALAH-DR3 releases with the updated stoichiometric model from BB20. We also did a propagation error study to evaluate the potential errors on the predicted PBB composition due to the observational spectroscopic uncertainties. The propagation error study consolidates the idea that the numerical mass fraction values are uncertain. We emphasise that detailed stellar abundances of planet host stars are needed to further understand the observed planet population. However, from the analysis of both surveys, common and robust global trends appear.

Here, we list the main results obtained for PBB composition:

\begin{itemize}

        \item \textsf{Metallicity dependence:} The first part of this work (Sect. \ref{SecMassFractionMetallicity}) follows the approach used by BB20. It focuses on the metallicity dependance of the PBB composition. This is done for APOGEE-DR17, GALAH-DR3, and GALAH-DR2. The Fe-bearing molecules (FeS, Fe$_3$O$_4$, and Fe$_2$O$_3$) show very stable trends as a function of [Fe/H], while the Mg-bearing molecules (MgSiO$_3$, Mg$_2$SiO$_4$) show a different behaviour in APOGEE-DR17 and GALAH-DR3. In APOGEE-DR17, the mass fraction of Mg$_2$SiO$_4$ and MgSiO$_3$ decreases and increases with [Fe/H], respectively, while the opposite is found for GALAH-DR3. Moreover, we have a common trend in both surveys since the MgSiO$_3$ mass fraction is always higher than Mg$_2$SiO$_4$. However, the PBB composition obtained with the GALAH-DR2 is inconsistent with the others surveys. The MgSiO$_3$ mass fraction is lower than the one of Mg$_2$SiO$_4$ (consistently with the results found by BB20 with GALAH-DR2). This discrepancy motivated a propagation error study. \\

        \item \textsf{Propagation error study:} With an error propagation calculation on the GALAH survey, we studied the sensitivity of the predicted mass fraction to the spectroscopic uncertainties. Given the typical error bars, the resulting mass fractions are largely impacted (\(\Delta m_{MgSiO_3}$$=$$\pm7\%\), \(\Delta m_{Mg_2SiO_4}$$=$$\pm8\%\), \(\Delta m_{H_2O}$$=$$\pm15\%\)), showing that the numerical mass fraction values are uncertain. However, unless used to consider extreme limit error cases, the main trends with metallicity [Fe/H] and the alpha content [$\alpha$/Fe] are preserved. In this sense, the double analysis on APOGEE and GALAH supports these results. In spite of the potential inhomogeneous methods to determine spectroscopic abundances, both surveys qualitatively reveal similar mass fraction trends. This suggests that the apparent trends with metallicity and alpha content maybe interpreted as robust enough across the stellar populations of our Galaxy.\\
                
        \item \textsf{Alpha content dependence:} One goal of our study is the analysis of the PBB composition in the [$\alpha$/Fe]-[Fe/H] plane, which enables us to compare galactic populations. We found an explicit mass fraction dependence of all species (FeS, Fe2O3, Fe3O4, Mg2SiO4, MgSiO3) with the alpha content [$\alpha$/Fe] of the stars. In case of water, the trend with [$\alpha$/Fe] is weaker than for other molecules and appears to be anti-correlated with the metallicity [Fe/H].\\
                        
        \item \textsf{Water mass fraction:} The stoichiometric relations are a simple way to predict the PBB composition. However, this method is maybe simplistic in case of water due to the ice-line presence. Indeed, the multiple processes happening in the proto-planetary disc can lead to very different planet formation scenarios with different water mass fractions in a planet's composition \citep[e.g.][]{Bitsch2021}. Furthermore, in this study we found that the water mass is more dependent on the metallicity that on the alpha content. The PBB water content is clearly anti-correlated with [Fe/H] for both surveys. However, there is no clear $\alpha$ dependence with the PBB water content. Because of a different C/O dependance with [Fe/H] and [$\alpha$/Fe], the APOGEE DR17 sample shows that the thin disc is more water-rich than the thick disc while the opposite is found for GALAH DR3 (Fig. \ref{Fig_Mfrac_ThinThickperFeH}). Whether the water abundance of PBB is a good proxy for planets or whether the thick disc stars (which are rather alpha-rich, metal-poor stars) are suitable hosts water worlds and/or hycean planets \citep{Madhusudhan2021} remains a matter of debate.\\

        \item \textsf{Bimodal PBB composition:} The dip of stellar occurrence density in the [$\alpha$/Fe]-[Fe/H] plane is usually interpreted as the separation of the thin and the thick discs. Our results with APOGEE and GALAH large surveys show that most of species have an $\alpha$-content dependence. As a consequence, the thin/thick dichotomy leads to a bimodal distribution of PBB. Interestingly, this has also been found by S17 with the HARPS survey and C19 with a stellar population synthesis model. This implies that the chemical composition in the early phases of proto-planetary discs could largely differ depending on the galactic origin of the host star. \\   

        \item \textsf{Interstellar objects:} In a similar fashion, we can expect that water-rich small bodies are more likely formed around metal-poor stars. The early history of planetary systems involves a period when the bulk of these bodies are ejected out of these systems. Therefore, the observation of interstellar objects when they reach our Solar System may provide clues as to the formation process of stars and planets, provided they are not too significantly modified after their formation. 
                
\end{itemize}

Despite the large amount of high-quality data, it appears obvious that the observational uncertainties are still large and make the analysis difficult. This work stresses the need for large surveys of very high spectroscopic quality. In particular, the oxygen and carbon measures are known to present large errors, which are crucial to determining robust PBB composition. 
However, while the propagation study reveals that the PBB composition is uncertain, the double study on APOGEE and GALAH suggests that the overall trends with metallicity and $\alpha$-content are robust for most of the molecules.

These results open a larger discussion on the impact of the chemical evolution of the Milky Way on the current planet populations. The potential role of icy PBB in planet formation theory should be investigated more deeply in a galactic context taking into account for the diversity of initial chemical abundances. Our results clearly indicate that the host star abundance can have significant influences on the proto-planetary discs since the PBB water content appears to be anti-correlated with metallicity. The potential impact of this anti-correlation on the planetary evolution should be investigated in details to fully understand the surveys of exoplanet observations.

Finally, a complete discussion on PBB composition across stellar populations and their translation into planets and small body properties should mention the potential presence of short-lived radiogenic nuclides such as ${}^{60}$Fe or ${}^{26}$Al, which have a strong potential to dehydrate planetesimals \citep{Lichtenberg2019NatAs} and modify their refractory composition due to liquid-rock interactions. The nuclide ${}^{26}$Al is produced by massive stars \cite[see e.g.][]{Forbes2021NatAs}, supernovae, or Wolf-Rayets, which are more frequent in the thin disc because it is younger than the thick one. In this sense, the potential presence of ${}^{26}$Al in the thin disk could decrease the final water mass fraction of bodies around thin disc stars. This effect could  increase the difference of PBB water content predicted around the thin and thick disc stars.

\begin{acknowledgements}
We thank an anonymous referee for many constructive comments, which greatly improved the quality and clarity of our
paper. Moreover, we thank Sven Buder for useful discussions. This project has received funding from the European Research Council (ERC) under the European Union’s Horizon 2020 research and innovation programme (Grant Agreement No 802699). B.B., acknowledges the support of the European Research Council (ERC Starting Grant 757448-PAMDORA) and of the DFG priority program SPP 1992 “Exploring the Diversity of Extrasolar Planets (BI 1880/3-1). N.L. acknowledges financial support from "Programme National de Physique Stellaire" (PNPS) and from the "Programme National Cosmology et Galaxies (PNCG)" of CNRS/INSU, France. 
\end{acknowledgements}

\bibliographystyle{aa}
\bibliography{reference}

\begin{appendix}

\section{Propagation error study}
\label{AppendixErr}

\begin{figure}[ht]
  \centering
   \includegraphics[width=6.9cm,clip=true, trim= 0cm 0cm 0cm 0cm]{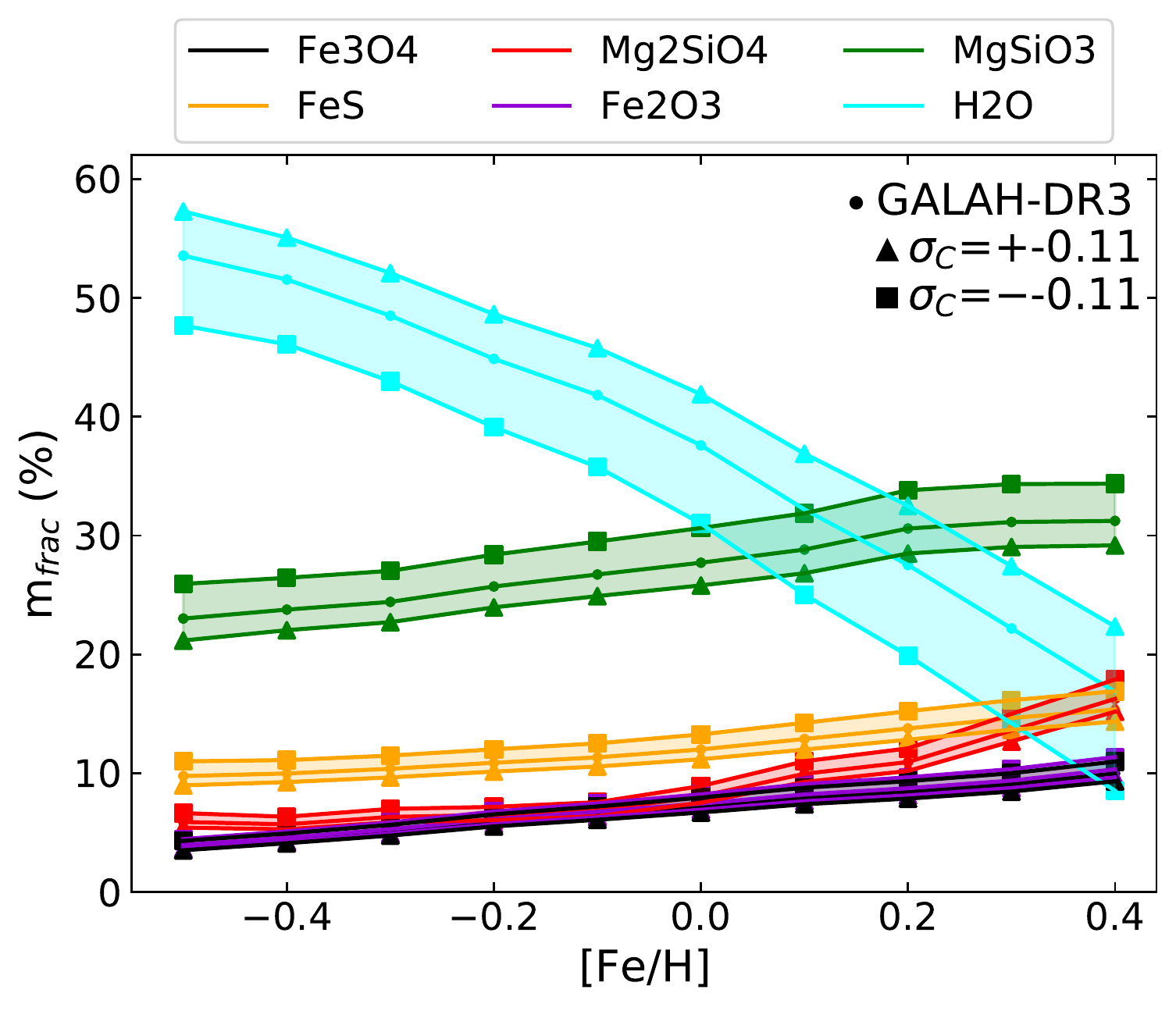}
   \includegraphics[width=6.9cm,clip=true, trim= 0cm 0cm 0cm 0cm]{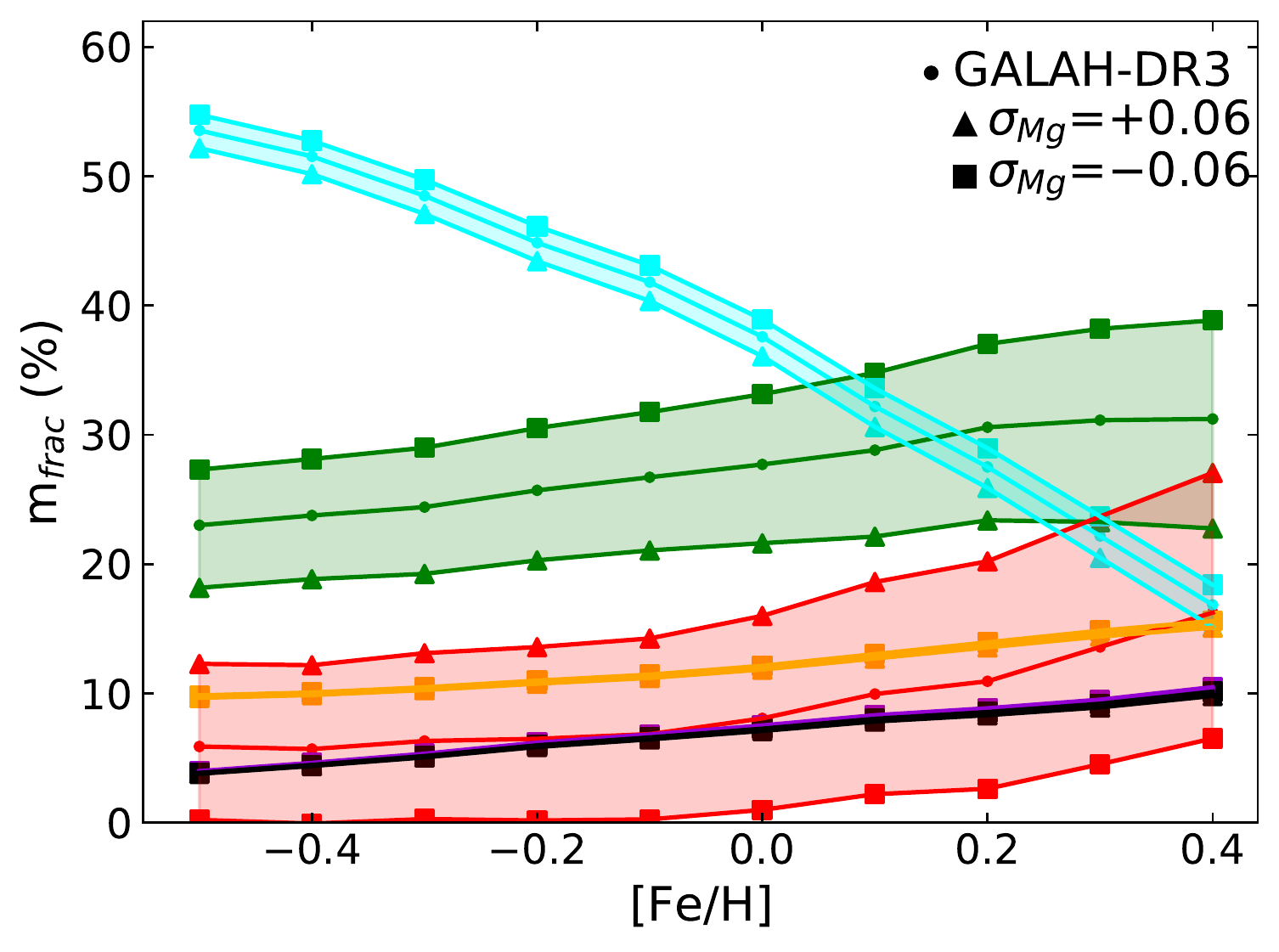}
   \includegraphics[width=6.9cm,clip=true, trim= 0cm 0cm 0cm 0cm]{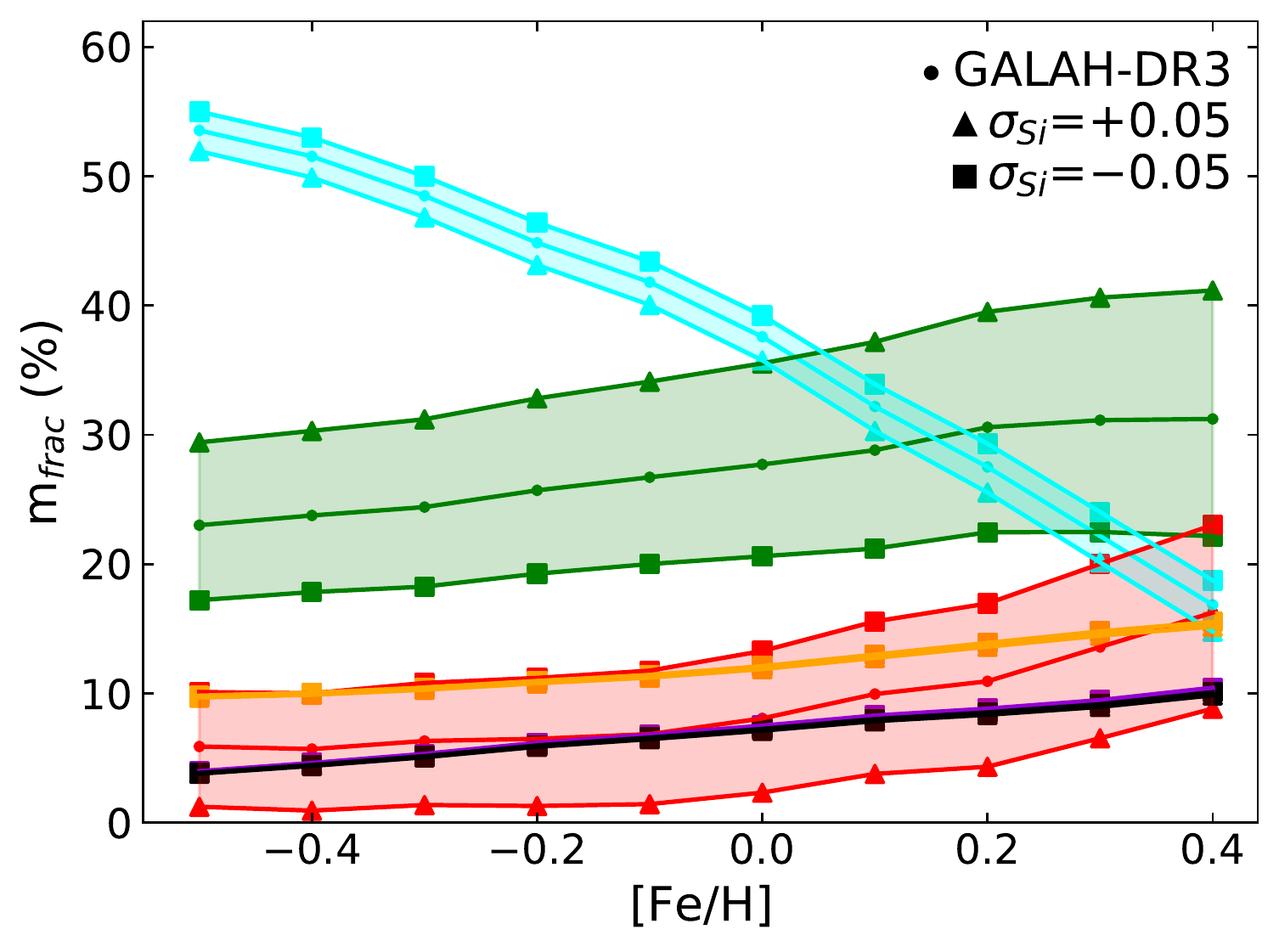}
   \includegraphics[width=6.9cm,clip=true, trim= 0cm 0cm 0cm 0cm]{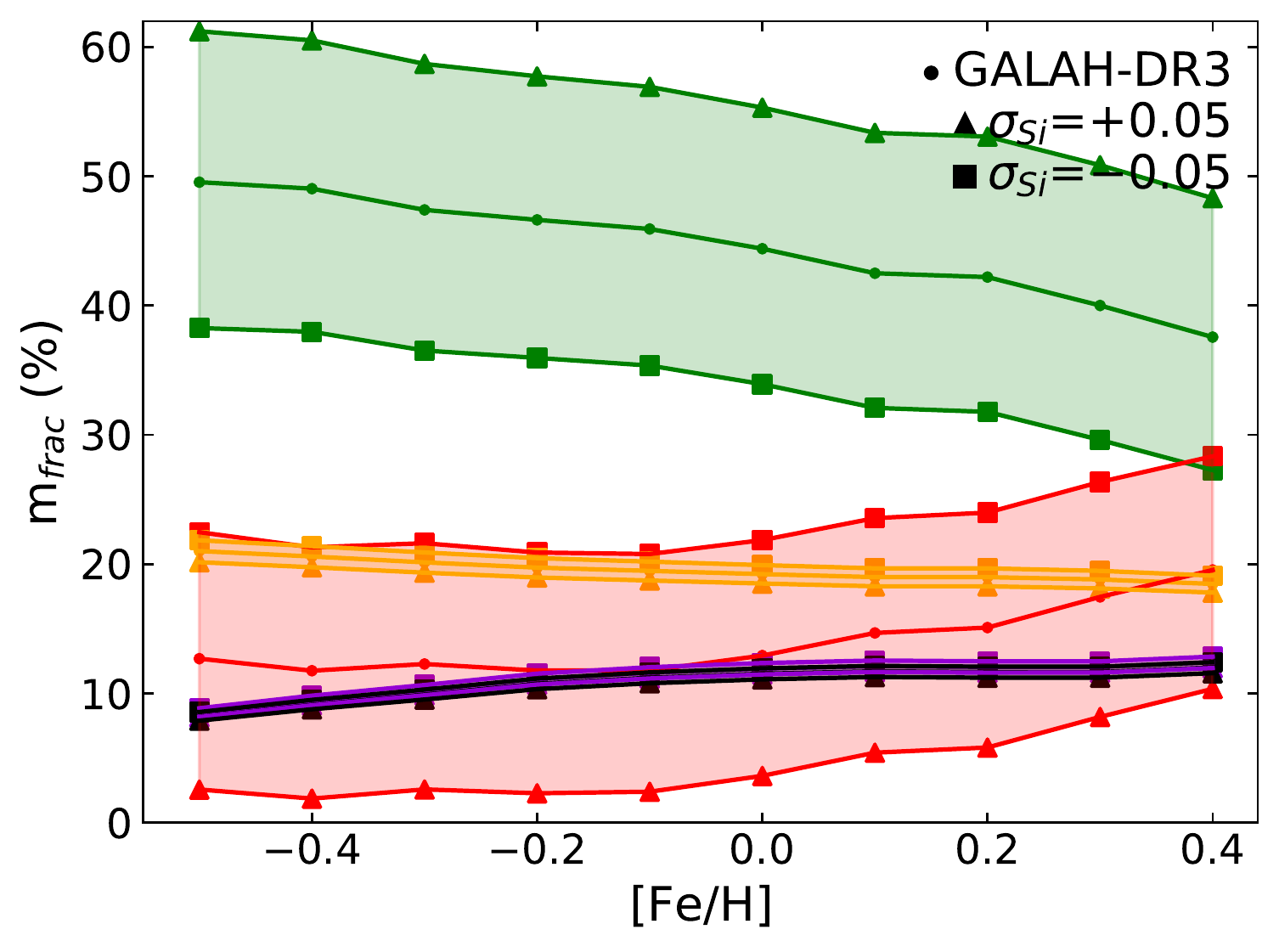}   
     \caption{Mean PBB mass fractions per bin of metallicity for GALAH-DR3. Panels consider T<150 K varying respectively $\sigma_C$, $\sigma_{Mg}$ and $\sigma_{Si}$. Bottom panel considers T>150 K varying $\sigma_{Si}$.}
 \label{FigAppendix}
\end{figure}

\section{Mean PBB mass fraction in the thin and the thick discs}
\label{AppendixNumberStarsGalactic}

\begin{table}[h]
\centering                                      
\begin{tabular}{c|cccc}          
                                &                               \multicolumn{4}{c}{GALAH DR3}                              \\

                                &      Thin                & Thick      & Intermediate   &        Halo   \\
\hline                        
\hline                        
                                &                               \multicolumn{4}{c}{Inner proto-planetary disk (T>150 K)}                           \\

\multicolumn{1}{c|}{ <$m_{\rm FeS}$> }              & 19                &  20              &  19     & -\\
\multicolumn{1}{c|}{ <$m_{\rm Fe_2O_3}$> }     & 12             & 10             &  11    & -\\ 
\multicolumn{1}{c|}{ <$m_{\rm Fe_3O_4}$> }     & 11             &  9             &  11    & -\\ 
\multicolumn{1}{c|}{ <$m_{\rm Mg_2SiO_4}$> } & 14               &  22            &  18   & -\\ 
\multicolumn{1}{c|}{ <$m_{\rm MgSiO_3}$> }     & 44             &  39         &  41   & -\\ 
\hline                                   

                                &                               \multicolumn{4}{c}{Outer proto-planetary disk (T<150 K)}                           \\
                                
\multicolumn{1}{c|}{ <$m_{\rm FeS}$> }              &  12               & 11               &  12   & -\\
\multicolumn{1}{c|}{ <$m_{\rm Fe_2O_3}$> }     &  8                     & 6                &  7  & -\\
\multicolumn{1}{c|}{ <$m_{\rm Fe_3O_4}$> }     &  7                     & 6                &  7  & -\\
\multicolumn{1}{c|}{ <$m_{\rm Mg_2SiO_4}$> } &  9                       & 13               & 11  & -\\
\multicolumn{1}{c|}{ <$m_{\rm MgSiO_3}$> }     &  27            & 22             &  25   & -\\
\multicolumn{1}{c|}{ <$m_{\rm H_2O}$> }             & 37                & 42               & 38    & -\\
\multicolumn{0}{}{}  \\

                                &                               \multicolumn{4}{c}{APOGEE DR17}                             \\
                                &             Thin                 & Thick         & Intermediate   &      Halo   \\
\hline                        
\hline                       
                                &                               \multicolumn{4}{c}{Inner proto-planetary disk (T>150 K)}                           \\

\multicolumn{1}{c|}{ <$m_{\rm FeS}$> }              &  18               &  18              &  18     & 20  \\
\multicolumn{1}{c|}{ <$m_{\rm Fe_2O_3}$> }     &  11            & 9              &   10   & 9 \\
\multicolumn{1}{c|}{ <$m_{\rm Fe_3O_4}$> }     &  11            &  9             &  10    & 9\\
\multicolumn{1}{c|}{ <$m_{\rm Mg_2SiO_4}$> } &  11              &  23            &  16   & 17 \\
\multicolumn{1}{c|}{ <$m_{\rm MgSiO_3}$> }      & 49            &  41         &  46   & 45\\
\hline                               

                                &                               \multicolumn{4}{c}{Outer proto-planetary disk (T<150 K)}                           \\

\multicolumn{1}{c|}{ <$m_{\rm FeS}$> }             & 10         & 10         &   10  & 11\\
\multicolumn{1}{c|}{<$m_{\rm Fe_2O_3}$>  }     & 7              & 5              & 7    & 6\\
\multicolumn{1}{c|}{ <$m_{\rm Fe_3O_4}$>  }    & 6              & 6              &   6  & 5\\
\multicolumn{1}{c|}{ <$m_{\rm Mg_2SiO_4}$> } & 7                & 13             &   9  & 9\\
\multicolumn{1}{c|}{ <$m_{\rm MgSiO_3}$> }     &  28    & 24             &   27   & 25\\
\multicolumn{1}{c|}{<$m_{\rm H_2O}$>  }          &  42  & 42             &   41  & 44\\

   \hline  
\multicolumn{0}{}{}  \\
      
\end{tabular}
\caption{Mean PBB mass fractions for Galactic components classified by the kinematical approach, for the internal disc (T>150 K) and the external disc (T<150 K). The selected GALAH DR3 sample have only one halo star (cf. Table \ref{Table1_GalComponents}), such that the mass fractions are not included.}
\label{Table3_MeanMfracCompGal}
\end{table}

\FloatBarrier

\begin{figure*}
 \centering
    \includegraphics[width=7cm,clip=true, trim= 0cm 0cm 0cm 0cm]{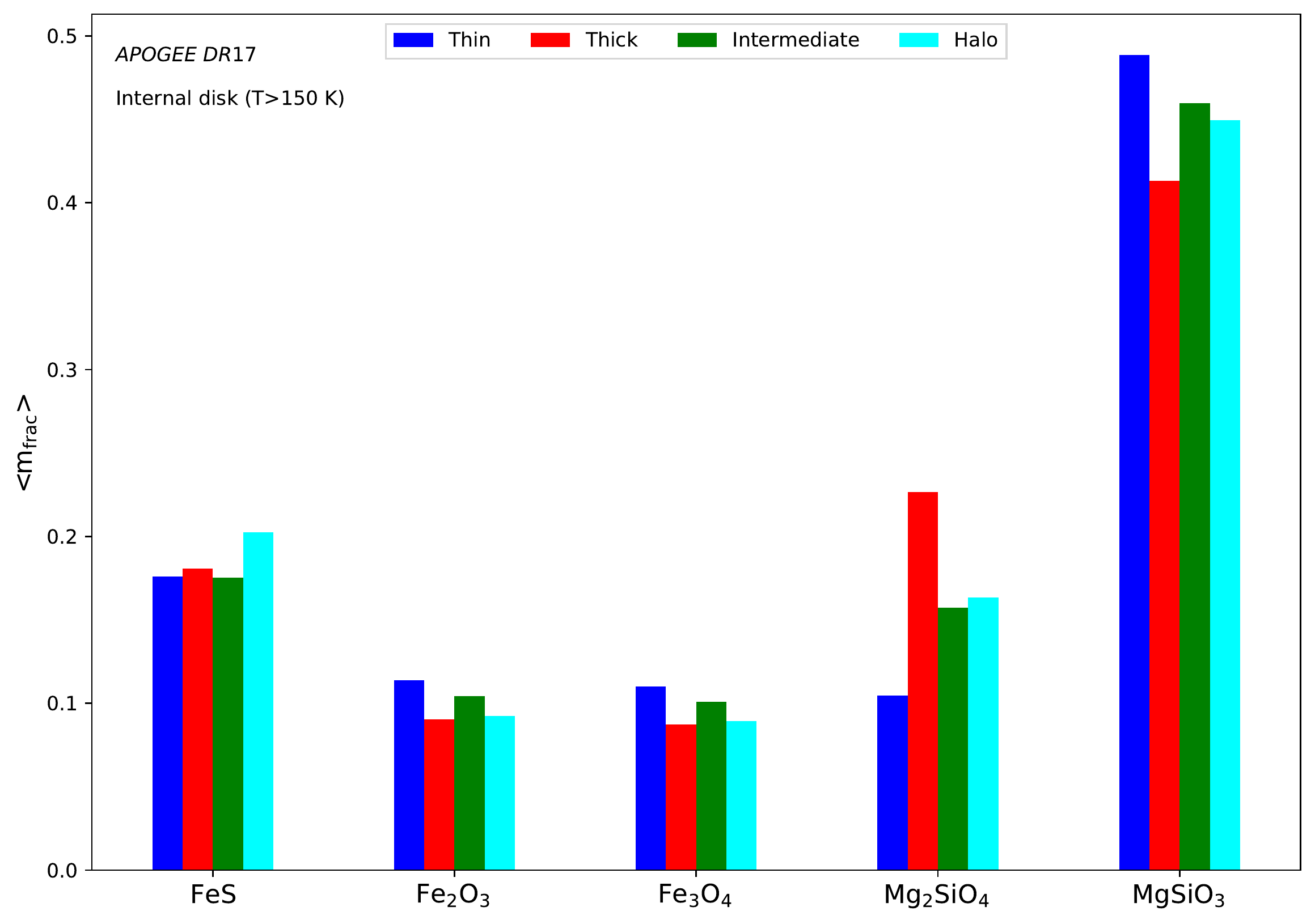}
    \includegraphics[width=7cm,clip=true, trim= 0cm 0cm 0cm 0cm]{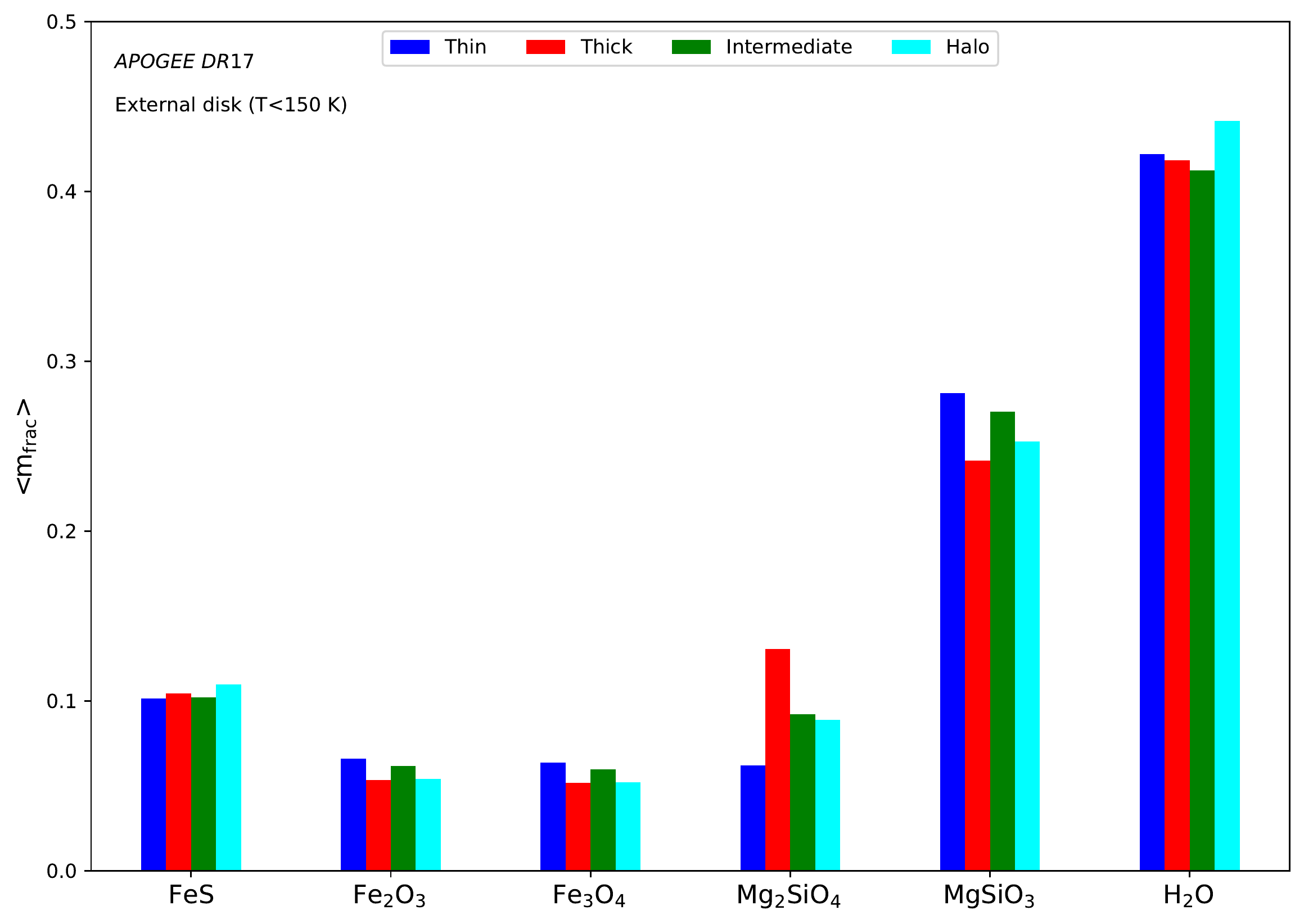}
   \includegraphics[width=7cm,clip=true, trim= 0cm 0cm 0cm 0cm]{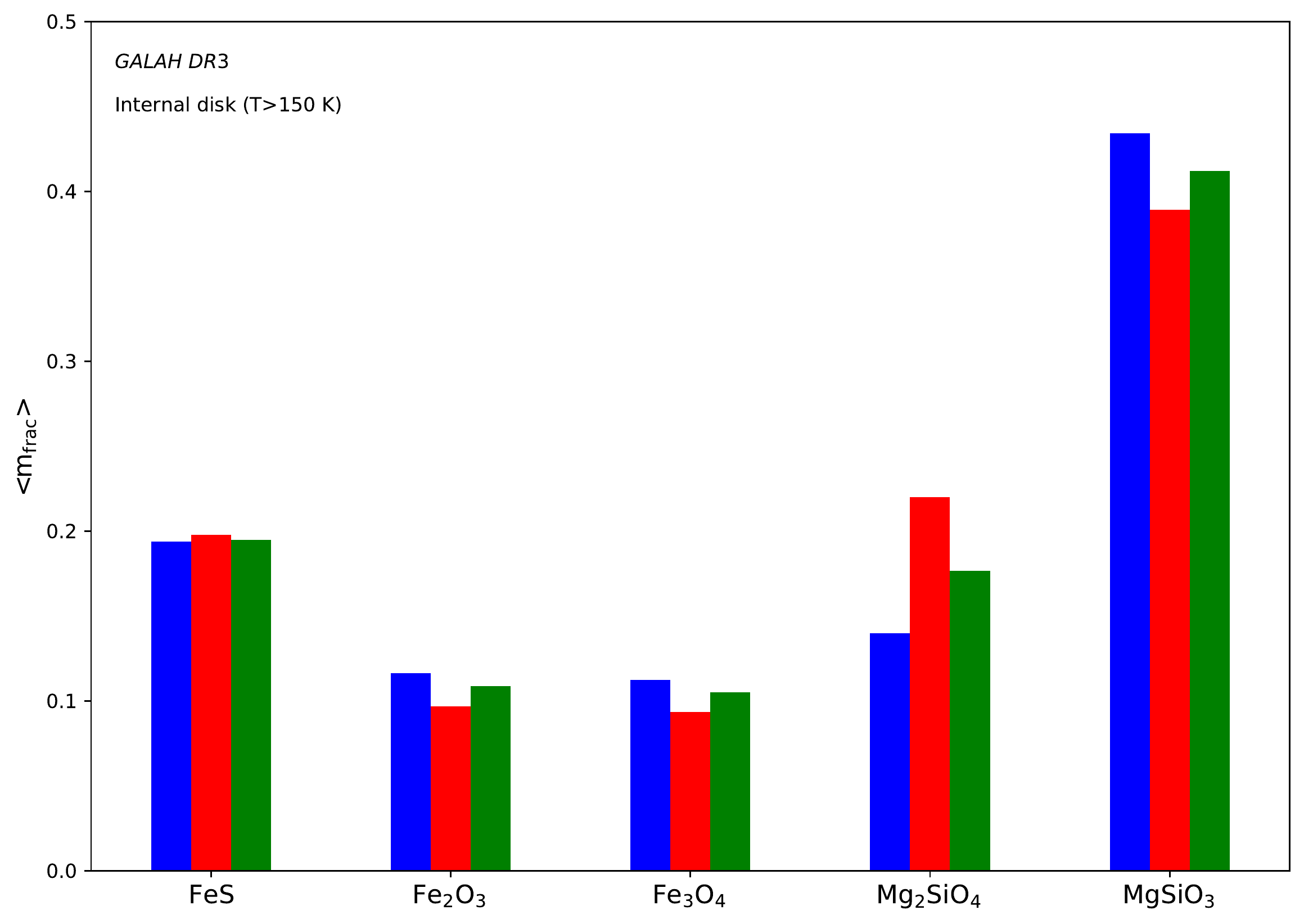}
   \includegraphics[width=7cm,clip=true, trim= 0cm 0cm 0cm 0cm]{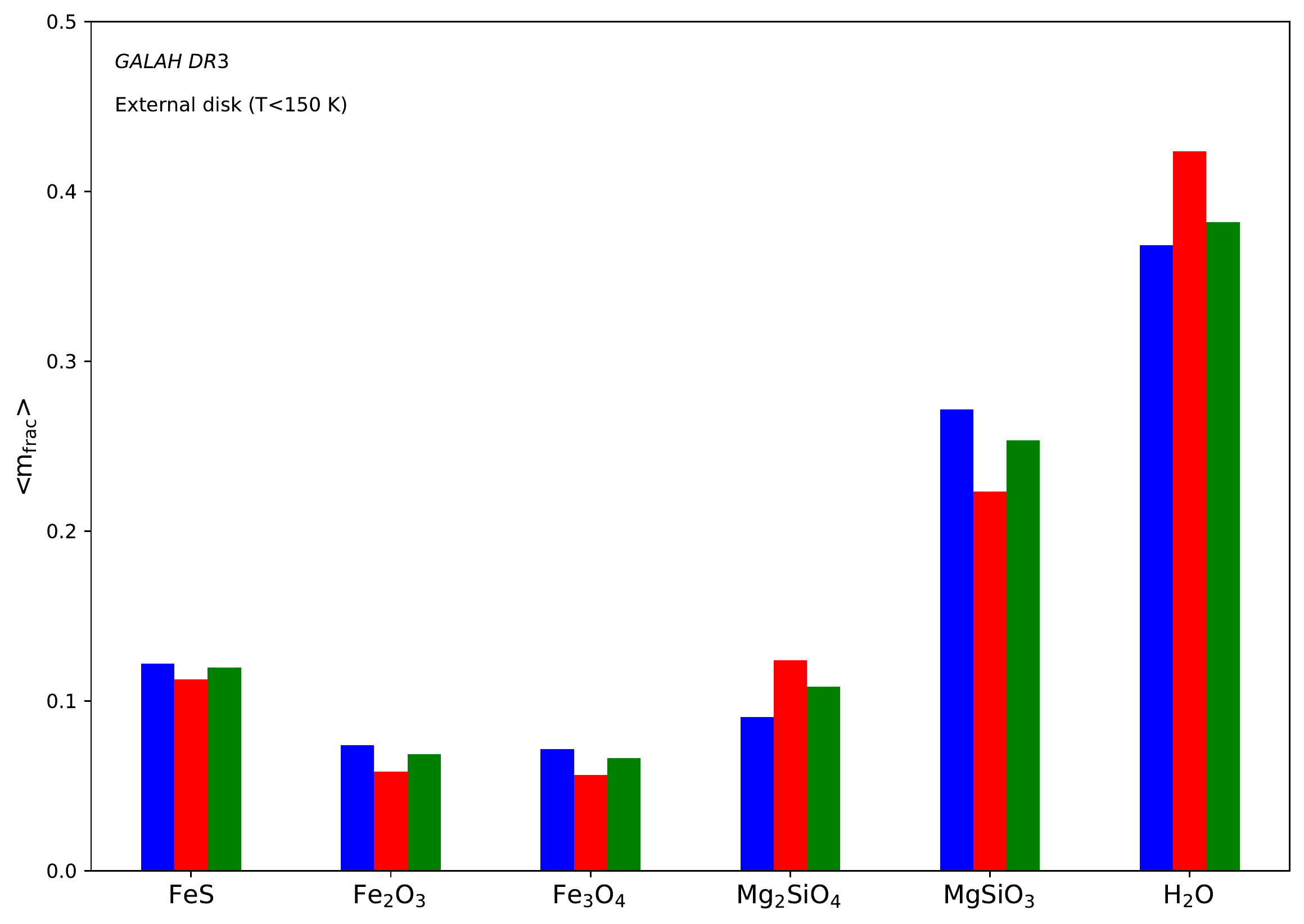}
     \caption{Mean PBB mass fractions for galactic components, the internal disc (T>150 K, left panels), and the external disc (T<150 K, right panels).  The selected GALAH DR3 sample only has one halo star, such that the mass fractions are not included.}
    \label{FigAppendixHistoKine}
\end{figure*}

\FloatBarrier

\section{Chemical classification}
\label{AppendixDensityPlot}

\begin{figure}

 \centering
    \includegraphics[width=6.3cm,clip=true, trim= 0cm 0cm 0cm 0cm]{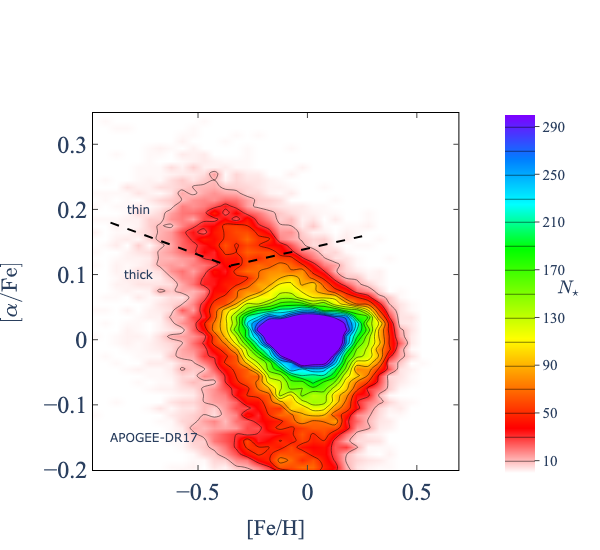}
    \includegraphics[width=6.3cm,clip=true, trim= 0cm 0cm 0cm 0cm]{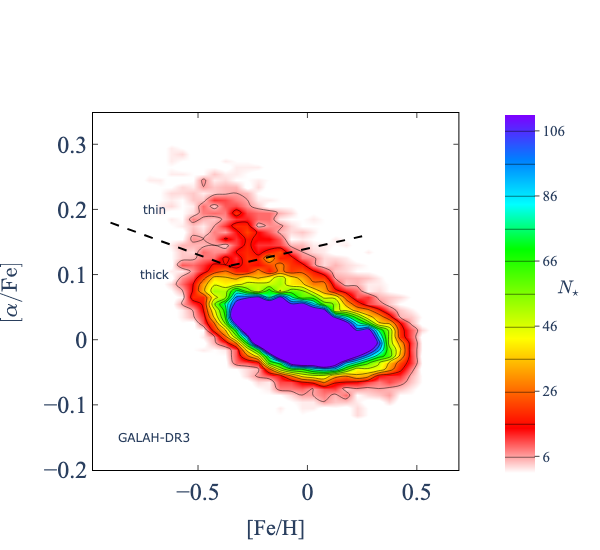}

     \caption{Stellar density is colour-coded in the [$\alpha$/Fe]-[Fe/H] diagram. The stellar density is standardly used to separate the thick disc from the thick disc, as is done with the dashed line. Top panel: APOGEE-DR17 sample; bottom panel: GALAH-DR3 sample.}
     \label{FigStellarDensity}
    
\end{figure}

In this appendix, we want to prove the robustness of our results by a different classification method. We apply here the standard chemical classification in the [$\alpha$/Fe]-[Fe/H] diagram taking into account the stellar density (colour-coded in Fig \ref{FigStellarDensity}). In this method, the thick disk is differentiated from the thin disc based on the stellar density. The following equations, separating the thin disc from the thick disc, are plotted in Fig. \ref{FigStellarDensity} as dashed lines:

 \begin{equation} 
         \begin{split}  
&\hspace*{0.3cm}  \hspace*{0.3cm}  [\alpha/Fe] = -0.123\times[Fe/H]+0.069 \hspace*{0.3cm}  \text{if [Fe/H]<-0.35:}    \\
&\hspace*{0.3cm}  \hspace*{0.3cm}  [\alpha/Fe] =  0.075\times[Fe/H]+0.14   \hspace*{0.7cm}  \text{if [Fe/H]>-0.35:.} 
         \end{split}
 \end{equation}

\begin{figure*}

 \centering
    \includegraphics[width=8cm,clip=true, trim= 0cm 0cm 0cm 0cm]{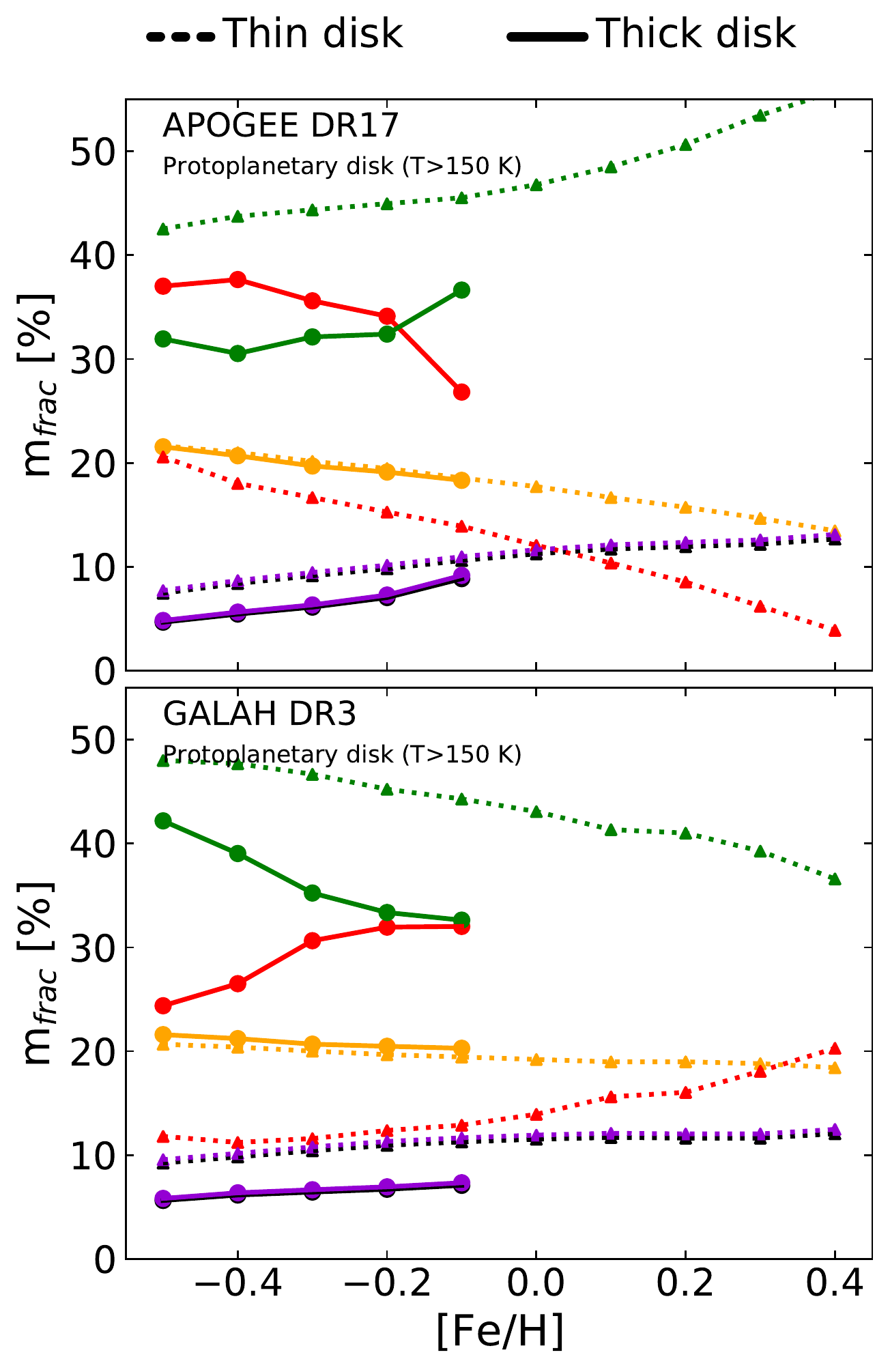}
    \includegraphics[width=8cm,clip=true, trim= 0cm 0cm 0cm 0cm]{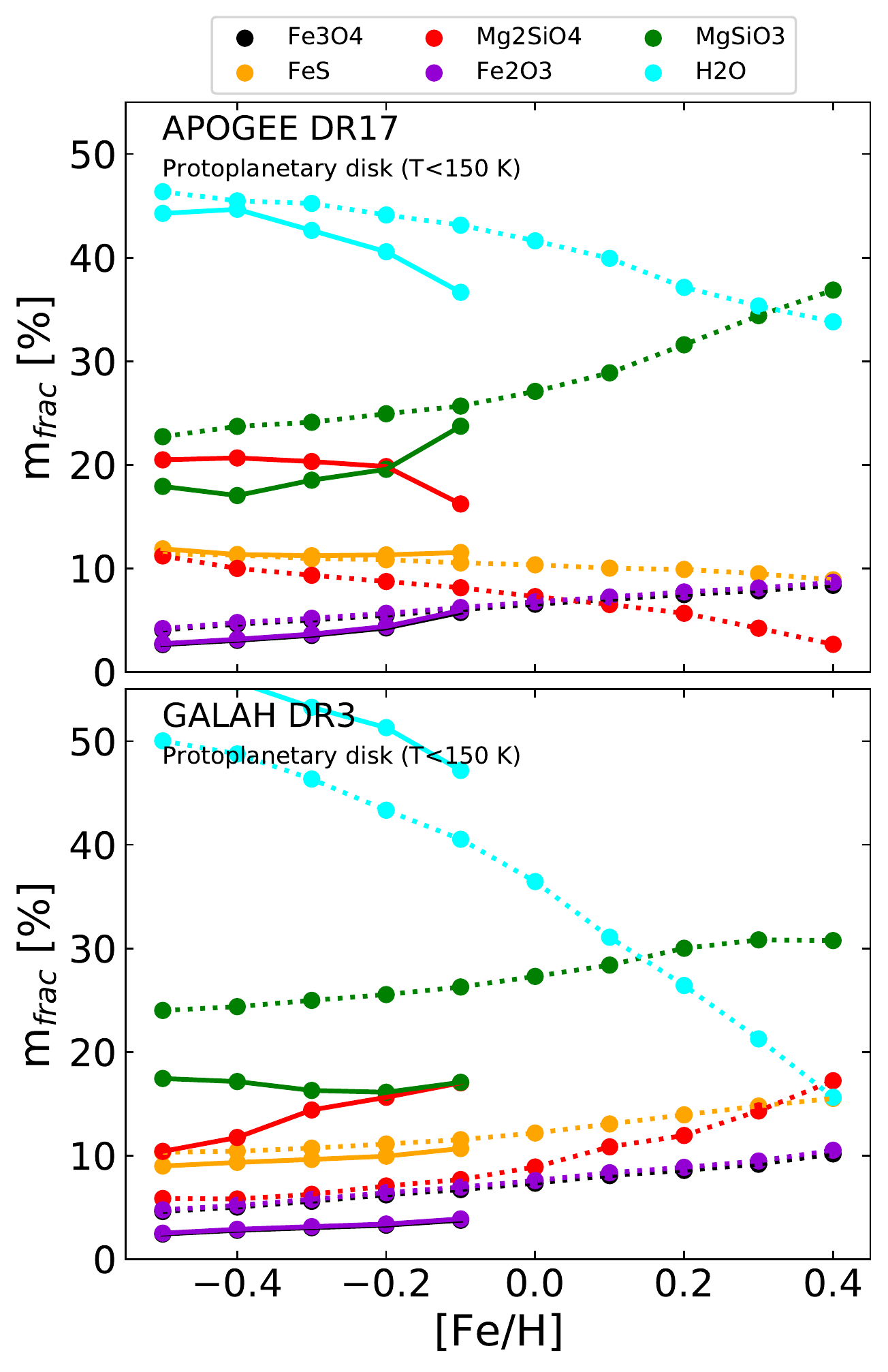}

     \caption{Same Figure as Fig. \ref{Fig_Mfrac_ThinThickperFeH}, but using the chemical classification method.}
     \label{FigMfracFeH_ChemicalClass}
    
\end{figure*}

Figure \ref{FigMfracFeH_ChemicalClass} presents some differences compared to Fig. \ref{Fig_Mfrac_ThinThickperFeH}. As expected, the different classification methods change the statistics of the thick disc and impact their mass fraction values. In the case of APOGEE-DR17, in both temperature cases (T<150 K and T>150 K) the Mg$_2$SiO$_4$ and the MgSiO$_3$ have reversed values in Fig. \ref{FigMfracFeH_ChemicalClass} and Fig. \ref{Fig_Mfrac_ThinThickperFeH}. However, with both classification methods we can see that the differences between the thick disc and the thin disc are similar: the Mg$_2$SiO$_4$ is higher in the thick disc than in the thin disc, and the opposite is found for the MgSiO$_3$. The other molecules also have very similar values in both classification methods. In particular, the water mass fraction trend is the same with both methods and the differences between the thin and the thick discs are kept equivalent. Finally, the thin disc mass fraction values are almost unchanged.

\FloatBarrier

\end{appendix}

\end{document}